\documentclass[a4paper,fleqn,usenatbib]{mnras}
\usepackage{graphicx}
\usepackage[T1]{fontenc}
\usepackage{ae,aecompl}

\usepackage{amsmath}	
\usepackage{amssymb}	

\usepackage{color}

\def\farcs{\hbox{$.\!\!^{\prime\prime}$}}
\newcommand{\kms}{km~s$^{-1}$\,}
\newcommand{\msun}{${\cal M}_\odot$\,}

\title[Compact multiple HIP 41431]{The compact multiple system HIP 41431}

\author[Borkovits et al.]{
T.~Borkovits$^{1,2}$\thanks{E-mail: borko@electra.bajaobs.hu}, 
J. Sperauskas$^{3}$\thanks{E-mail: julius.sperauskas@ff.vu.lt}, 
A.~Tokovinin$^{4}$\thanks{E-mail: atokovinin@ctio.noao.edu}, 
D. W. Latham$^{5}$\thanks{E-mail: dlatham@cfa.harvard.edu}, 
\newauthor
I. Cs\'anyi$^1$,
T. Hajdu$^{2,6,7}$,
L. Molnár$^{2,7}$\\
$^1$ Baja Astronomical Observatory of Szeged University, H-6500 Baja, Szegedi \'ut, Kt. 766, Hungary \\
$^2$ Konkoly Observatory, Research Centre for Astronomy and Earth Sciences, Hungarian Academy of Sciences, \\
 H-1121 Budapest, Konkoly Thege Miklós \'ut 15-17, Hungary \\
$^3$ Vilnius University Observatory, \v{C}iurlionio 29, 03100 Vilnius, Lithuania \\
$^4$ Cerro Tololo Inter-American Observatory, Casilla 603, La Serena, Chile\\
$^5$ Center for Astrophysics | Harvard \& Smithsonian, 60 Garden Street, Cambridge, MA 02138, USA \\
$^6$ E\"otv\"os Lor\'and University, Department of Astronomy, H-1118 Budapest, P\'azm\'any P\'eter stny. 1/A, Hungary\\
$^7$ MTA CSFK Lend\"ulet Near-Field Cosmology Research Group, H-1121, Budapest, Konkoly Thege Mikl\'os \'ut 15-17, Hungary\\
}

\begin{document}

\date{-}

\pagerange{\pageref{firstpage}--\pageref{lastpage}} \pubyear{2019}

\maketitle

\label{firstpage}

\begin{abstract}

The nearby (50  pc) K7V dwarf HIP~41431 (EPIC  212096658) is a compact
3-tier hierarchy.  Three K7V stars with similar masses, from  0.61 to 0.63
solar, make a triple-lined spectroscopic system where the inner binary
with a period  of 2.9 days is eclipsing, and the  outer companion on a
59-day  orbit  exerts  strong  dynamical influence  revealed  by  the
eclipse  time  variation  in  the {\em Kepler}  photometry.   Moreover,  the
centre-of-mass of the triple system moves on a 3.9-year orbit, modulating
the  proper motion.   The mass  of the  4-th star  is 0.35
solar.  The Kepler and ground-based photometry  and radial
velocities  from four  different  spectrographs are  used  to adjust  the
spectro-photodynamical  model  that accounts  for  dynamical  interaction.
The mutual inclination between the two inner orbits is 2\fdg16$\pm$0\fdg11, 
while the outer orbit is inclined to their common plane by 21\degr$\pm$16\degr. 
The inner orbit precesses under the influence of both outer orbits, 
causing observable variation of the eclipse depth.  Moreover, the phase 
of the inner binary is strongly modulated with a 59-day period and its line of apsides precesses. 
The  middle orbit with eccentricity $e=0.28$ also precesses, causing the observed variation of its radial velocity curve. 
Masses  and other  parameters of  stars in  this unique  hierarchy are
determined. This system is dynamically stable and likely old. 
\end{abstract}



\section{Introduction}
\label{sec:intro}

Study of stellar  hierachies containing three or more  bodies helps to
understand  their origin, still  a matter  of controversy  and debate.
Although the main  aspects of star formation are  well understood, the
genesis of  stellar systems,  particularly close binaries,  is obscure
because the  mechanisms reponsible for  bringing together two  or more
stars,  initially  formed  at   a  much  larger  separation,  are  not
identified or  modelled.  From  the observational side,  establishing a
reliable statistics of hierarchies  is a basis for testing theoretical
predictions.  However, individual  systems with rare
and/or  extreme  characteristics are  equally  enlightening, as  such
objects challenge the formation  theories and extend the boundaries of
the explored parameter space. This is the case under study here.

\begin{figure}
\centerline{
\includegraphics[width=8cm]{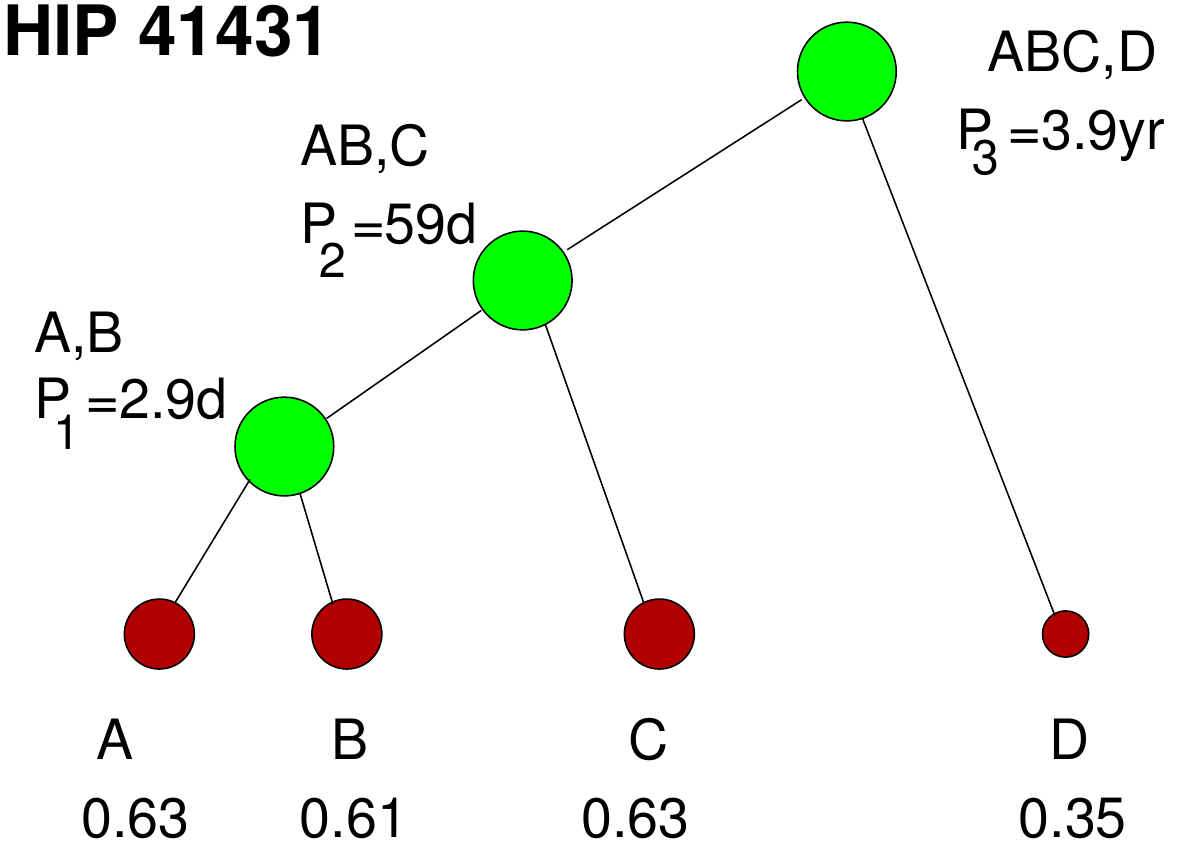}
}
\caption{Architecture of the quadruple system HIP~41431. Brown circles
  denote stars A to D, the numbers are their masses, green circles are
  subsystems. 
\label{fig:mobile} }
\end{figure}

We  investigate an interesting  low-mass hierarchical  stellar system,
HIP~41431 (GJ 307).   Basic data on this star  collected with the help
of Simbad are assembled in Table~\ref{tab:object}. This object came to
the attention  of the present authors independently  as a triple-lined
spectroscopic system (D.L.  and J.S.)  and as an  eclipsing binary, 
exhibiting fast and large amplitude eclipse timing variations (ETV) 
of likely dynamical origin (T.B. and T.H.).  Its architecture  is  illustrated in  Fig.~\ref{fig:mobile},
where the  three spectroscopically visible  components with comparable
masses and luminosities are designated as A, B, and C. The fourth star
D  was discovered by  the modulation  of radial  velocity (RV)  of the 
centre of  mass of the inner  triple and from the residuals of the dynamical, 
three-body ETV model, and confirmed  by its astrometric
signature in the  {\it Gaia} catalog.  All orbits seem  to be close to
one plane  and have small  eccentricities, resembling in this  sense a
solar   system,  like  the   ``planetary''  quadruple   star  HD~91962
\citep{planetary}. In these systems, the moderate period ratios on the
order  of  20 favor  dynamical  interaction  between  inner and  outer
orbits,  so  the  motion  cannot  be modelled  as  a  superposition  of
independent Keplerian orbits. However, compared  to HD~91962, HIP~41431 is much
more compact and fast.

The  paper   begins  with   a  short  description   of  the   data  in
Section~\ref{sec:obs}.   Then in  Section~\ref{sec:orb} we  present  and discuss
spectroscopic orbits and determine the preliminary components' masses.
Global  dynamical modelling of  the {\it  Kepler} K2  and ground-based
photometry and  the RVs is presented in Section~\ref{sec:dyn}. Its results
are confronted  with empirical  and theoretical stellar  properties in
\S~\ref{sec:par}.   Observed   effects   associated   with   dynamical
interaction  between the orbits  are covered  in Section~\ref{sec:orbprop}.
We summarize and discuss our findings in Section~\ref{sec:disc}.

\section{Observational data}
\label{sec:obs}

\begin{table}
\centering
\caption{Main characteristics of HIP 41431}
\label{tab:object}
\medskip
\begin{tabular}{l l  }
\hline
Parameter & Value \\
\hline
Identifiers & HIP~41431, GJ~307 \\
            &EPIC 212096658  \\
Position (J2000, Gaia DR2) &  08:27:00.91, +21:57:24.7 \\
PM $\mu_\alpha$, $\mu_\delta$ (mas~yr$^{-1}$, UCAC4) & +9.2$\pm$1.1, +21.08$\pm$1.1 \\
Parallax (mas, Gaia DR2) & 20.06 $\pm$ 0.09 \\
Spectral type & K7V \\
Optical photometry $B$, $V$, $G$ (mag) & 13.01, 10.84, 10.15 \\
Infrared photometry $J$, $H$, $K$ (mag) & 8.02, 7.37, 7.19 \\
Spatial velocity $U,V,W$ (\kms) & 8.1, 7.7, $-$1.4 \\ 
\hline
\end{tabular}
\end{table}

\subsection{Photometric observations}

\subsubsection{{\em Kepler K2} photometry}
\label{subsubsec:K2obs}

\begin{figure*}
\begin{center}
\includegraphics[width=0.49 \textwidth]{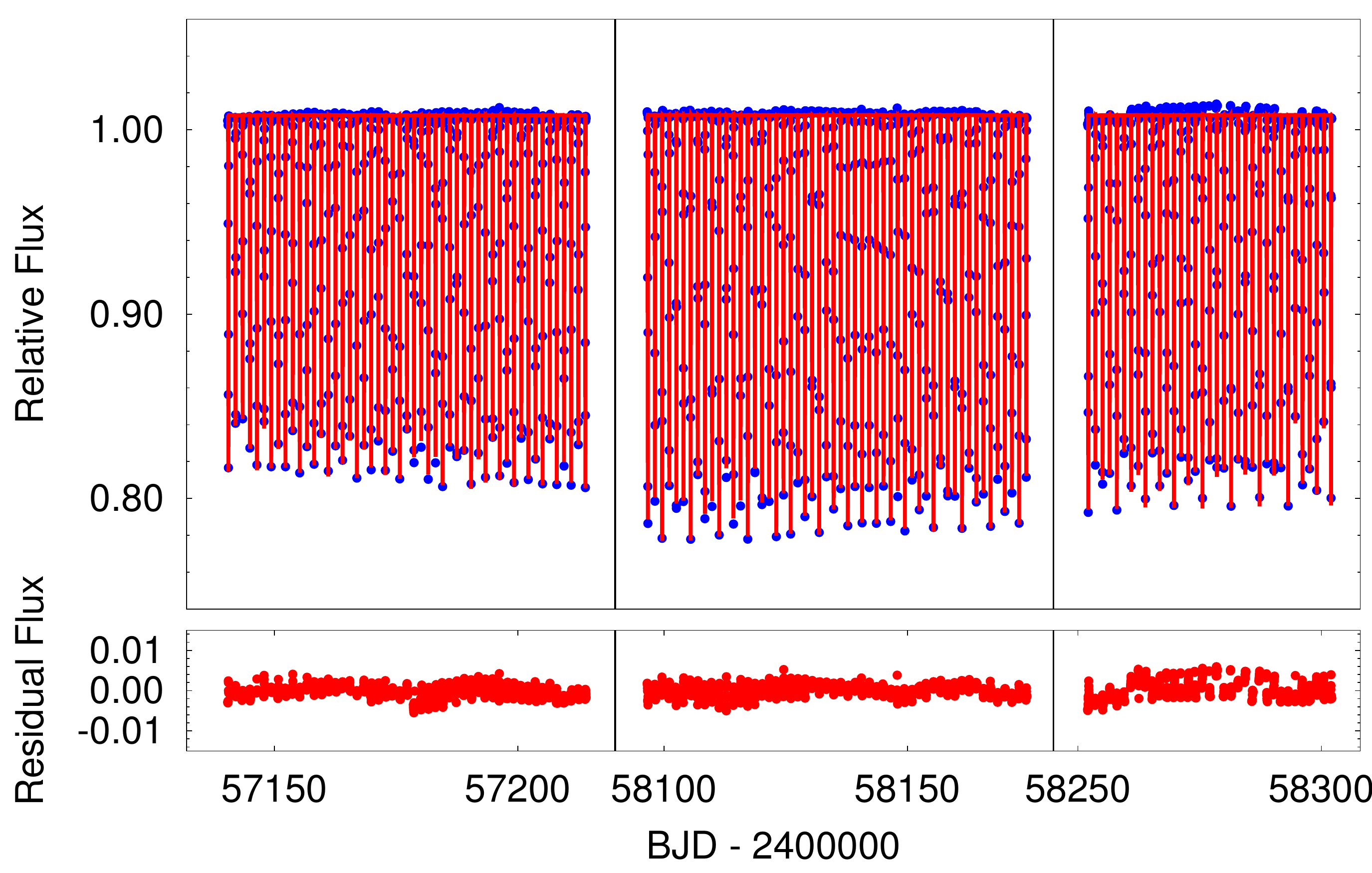}\includegraphics[width=0.49 \textwidth]{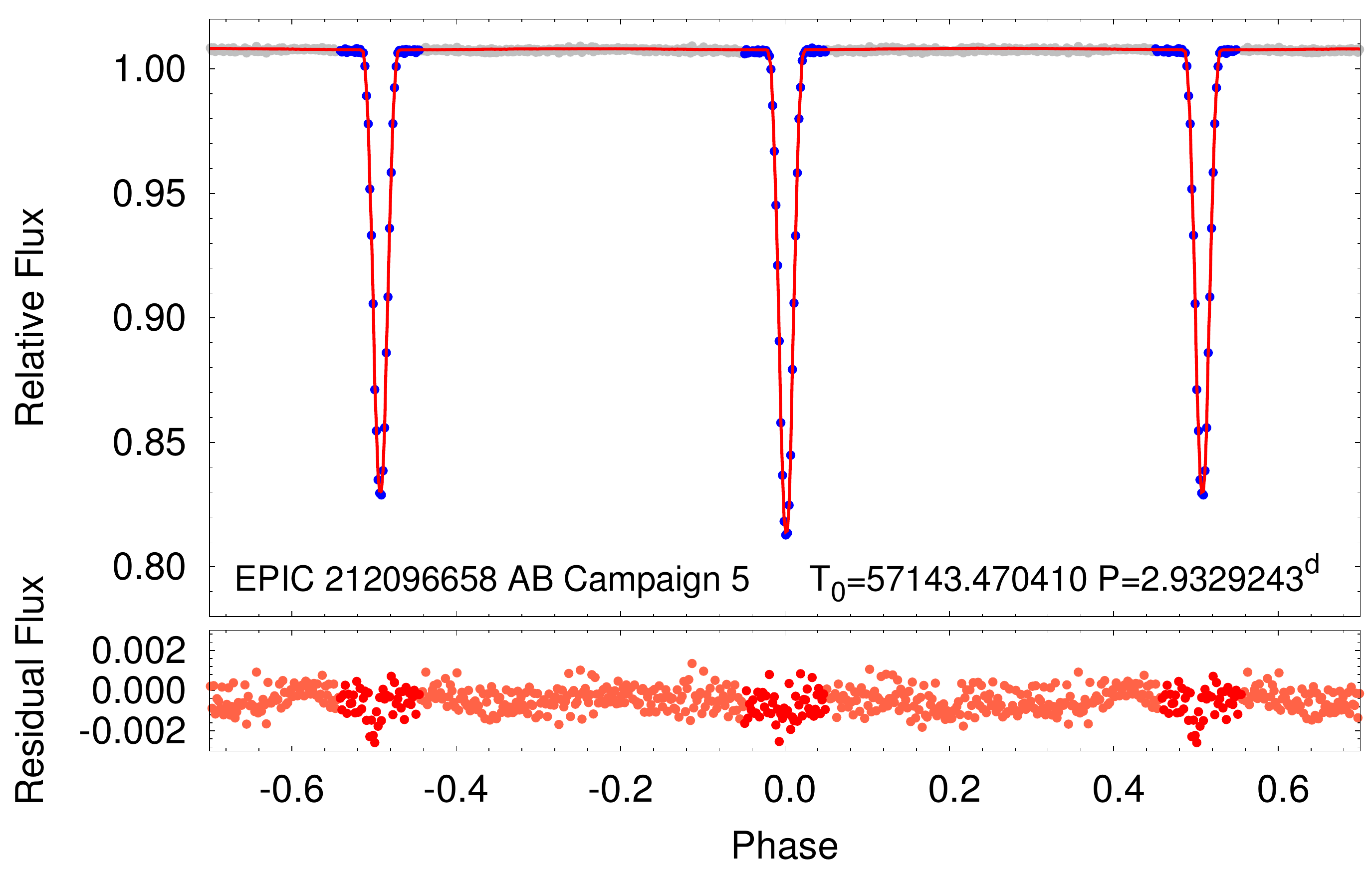}
\caption{ The {\em K2} lightcurves of HIP\,41431 (EPIC\,212096658). {\em Left panel}: The Campaigns 5, 16 and 18  long cadence PDCSAP lightcurves (blue circles) indicate moderate eclipse depth variations from campaign to campaign (see text for details). Red lines show our spectro-photodynamical model solution (Sect.\,\ref{sec:dyn}). {\em Right panel:} The phase-folded, binned, and averaged Campaign~5 {\em K2}-lightcurve of the innermost binary. The  phased averages of the observed flux near the eclipses are  plotted by the blue circles (these data were used for the joint spectro-photodynamical analysis), while the out-of-eclipse  flux is plotted by grey circles. The red curve is the folded, binned and averaged lightcurve of the cadence-time corrected photodynamical model solution calculated at the time of each observation; the residuals to the model are also shown in the bottom panels.}
\label{fig:eclipsefit} 
\end{center}
\end{figure*}

\begin{figure*}
\centerline{
\includegraphics[width=16cm]{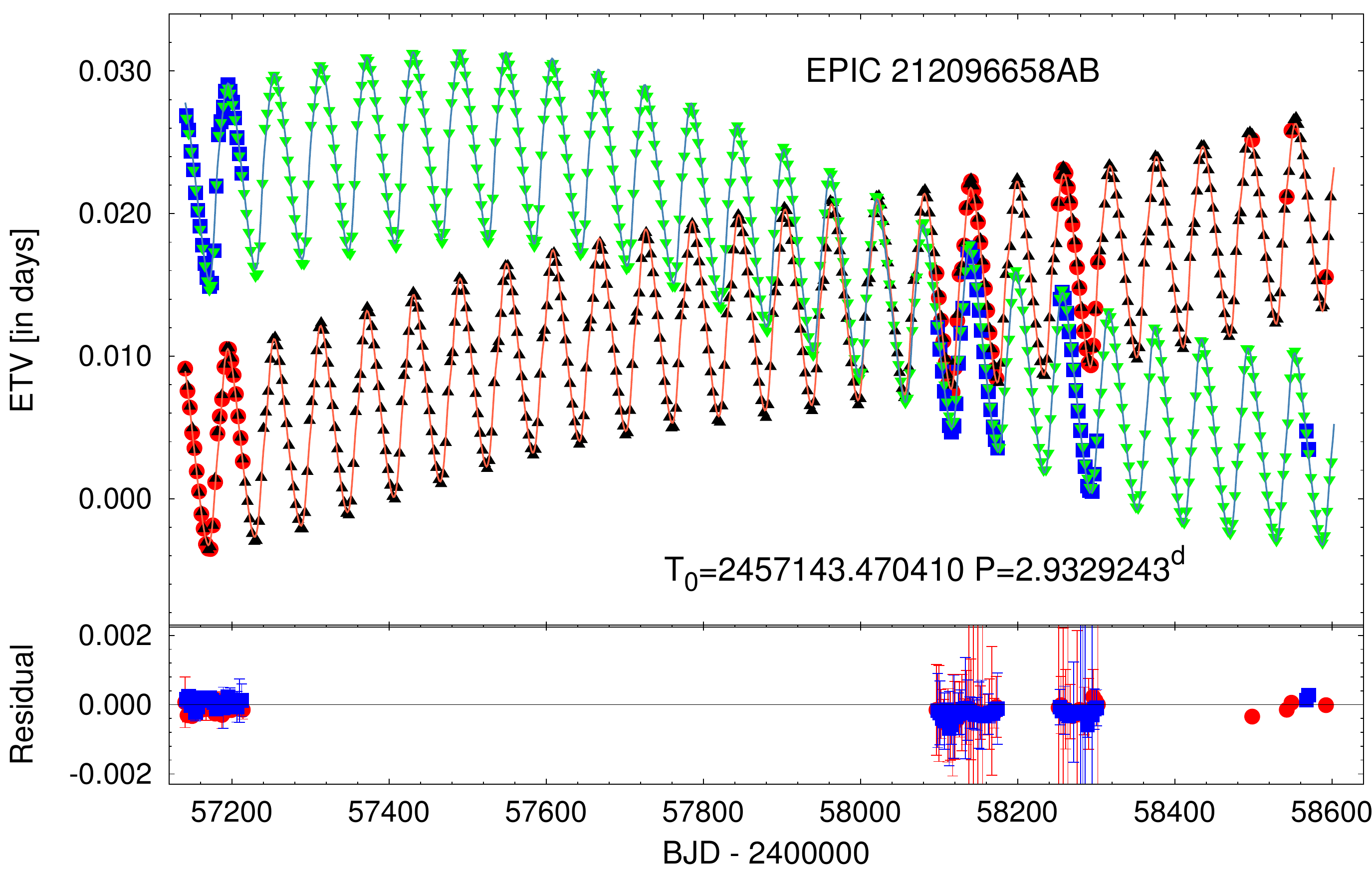}
}
\caption{ Eclipse timing variations of the innermost, eclipsing pair. Red circles and blue boxes stand for the primary and secondary ETVs, respectively, calculated from the observed eclipse events, while black upward and green downward triangles show the corresponding primary and secondary ETV, determined from the spectro-photodynamical model solution. Furthermore, orange and lightblue lines represent approximate analytical ETV models for the primary and secondary eclipses. The residuals of the observed vs photodynamically modelled ETVs are plotted in the bottom panel.}
\label{fig:ETV}
\end{figure*}

 HIP\,41431 was observed with {\em Kepler} spacecraft \citep{boruckietal10} in long cadence (LC) mode during Campaigns 5, 16 and 18 of {\em K2} mission. Furthermore, in Campaign 18 short cadence (SC) data were also collected. Eclipses with a period of 2.93291 days in the C5 data were reported by \citet{Barros2016}. Figure~\ref{fig:eclipsefit} shows the {\em K2} photometry and its model discussed below.

 We determined the mid-time of each observed eclipse and generated the ETV curves. The method we used is described in detail by \citet{Borkovits2016}. The times of minima are listed in Table\,\ref{Tab:EPIC_212096658_ToM}, while the ETV curves are shown in Fig.~\ref{fig:ETV}. The amplitude of the cyclic variation with  the 60-d period is 0.007 d, an order  of magnitude  larger than  the  light-time delay  in the  outer orbit. This variation is caused  primarily by the interaction with the star C that modulates the orbital elements of the inner orbit, including its period. Moreover, the inner orbit has a fast  apsidal rotation  which is also forced dynamically by star C. 

In such a compact,  strongly interacting triple system, even  marginally (by $\approx1-2\degr$) misaligned inner and outer orbital planes produce fast precession of the inner orbit and, hence, eclipse depth variations. Therefore, we checked the {\em K2} lightcurves for such features. Raw {\em K2} LC data are processed and corrected with different pipelines, resulting in somewhat different lightcurves. We downloaded from the  Barbara A. Mikulski Archive for Space Telescopes (MAST)\footnote{\url{http://archive.stsci.edu/k2/data_search/search.php}} and investigated the PDCSAP lightcurves obtained with the Kepler/K2 pipeline and the  {\em K2} self-flat-fielding ({\sc K2SFF}) pipeline of \citet{vanderburg2014}. Regarding the PDCSAP lightcurves, normalizing the flux levels of each of the three datasets to their out-of-eclipse averages reveals that the eclipses  in the C16 and C18 data are deeper by about $\approx6$\% and $1.5$\%, respectively,  relative to the C5 data. The same feature can be identified in the {\sc K2SFF} lightcurves, too. This finding, however, does not mean automatically that the  eclipse depth variation is real. Different locations of the target on the {\em Kepler}'s CCDs and different aperture masks used for the  photometry in these  datasets may produce apparent eclipse depth variation because of different amounts of contaminating fluxes from other stars within the apertures. However,  there are no stars brighter than $G=19.7$ mag within  $1.2'$ radius from our target in the {\it Gaia} DR2 catalog.  We conclude that slight eclipse depth variation during the observing window of {\em K2} photometry  is possible, although it cannot be proven conclusively. Therefore, we decided to apply our photodynamical modelling package both for the uneven and the uniform eclipse depth lightcurve. For the latter, we transformed the C5 and C18 lightcurves to have equal eclipse depths to the C16 data which exhibit the deepest eclipses.\footnote{In what follows, we will refer to those two kinds of lightcurves and the corresponding photo-dynamical solutions as the uneven and uniform eclipse depths scenarios.}

\subsubsection{Ground-based follow up photometry}

 In order to monitor the possible quick eclipse depth variations and  to lengthen the interval of the available ETV data suitable for the study of the dynamical evolution of the system, we carried out additional eclipse event observations with the 0.5 m telescope of Baja Astronomical Observatory of Szeged University located at Baja, Hungary, and equipped with an SBIG ST-6303 CCD detector. The target was observed on 7 nights between Jan 14 and Apr 18, 2019, which led to the determination of 6 additional times of minima data  (see also in Fig.\,\ref{fig:ETV} and Table\,\ref{Tab:EPIC_212096658_ToM}). The usual data reduction and photometric analysis were performed using {\sc IRAF}\footnote{{\sc IRAF} is distributed by the National Optical Astronomy Observatories, which are operated by the Association of Universities for Research in Astronomy, Inc., under cooperative agreement with the National Science Foundation.} routines.

As  shown below in Sect.\,\ref{sec:orbprop}, these observations confirmed not only the existence, but even the rate of the eclipse depth variations that was predicted by the uneven eclipse depth model solution.

\subsection{High-resolution spectroscopy}
\label{subsec:spectroscopy}

High-resolution spectroscopy was conducted independently using several
facilities. The primary goal was measurement of RVs for orbit
determination. Stellar parameters such as rotation, metallicity and
gravity can be determined as well from the spectra.  We tabulate the measured radial velocities in Table\,\ref{Tab:EPIC_212096658ABC_RV}.

\subsubsection{CfA observations}
\label{sec:CfA}


This  nearby  star was  observed  with  two identical  CfA Digital  Speedometers
\citep{Latham1985,Latham1992} from 1999.3  till 2008.3.  Seven observations were carried out with the instrument installed at the 1.5-m Wyeth Reflector at the Oak Ridge Observatory in  the town  of Harvard, Massachusetts. The other spectra were obtained with the 1.5-m Tillinghast Reflector at the Whipple Observatory on Mount Hopkins,  Arizona.  A total of 102  observations were collected.
The  RVs were  measured by  correlations of  the single  echelle order
centered on the Mg b triplet near 519\,nm, with a wavelength window of
4.5\,nm and resolving power of 35\,000. As  the spectrum is triple-lined, the  {\tt TRICOR} algorithm
was used,  analogous to {\tt TODCOR} \citep{TODCOR}.   A correction of
+0.14  \kms  must be  added  to  these RVs  to  put  them  on the  IAU
system. D.L. found the flux ratio C:A:B of 1:0.94:0.78 at 5187\AA. 

In 2009,  the  new fibre-fed  Tillinghast Reflector  Echelle
Spectrograph  \citep[TRES;][]{TRES} was used  to obtain  an additional
spectrum, followed by five more spectra taken in 2014. We measured the
RVs  by  cross-correlating these  spectra  with  the  binary mask  and
applied the zero-point correction of $-0.62$ \kms appropriate for this
instrument.

\subsubsection{VUES observations}

\begin{figure}
\centerline{
\includegraphics[width=8.5cm]{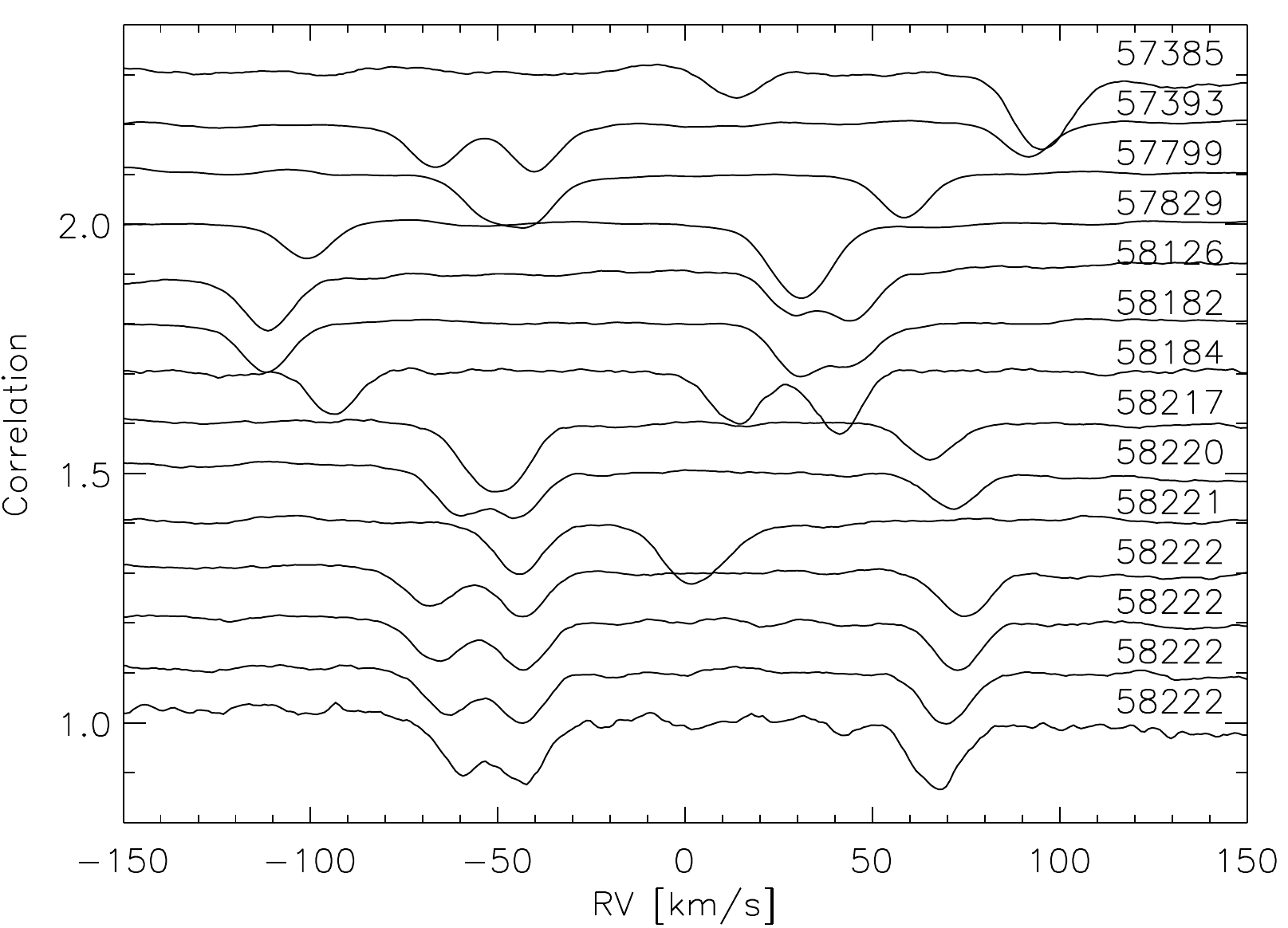}
}
\caption{CCFs  of  HIP~41431   recorded  with  VUES,  with  vertical
  shifts. The reduced Julian dates are indicated on the right.
\label{fig:VUES} }
\end{figure}


One of the  authors (J.S.)  has been conducting  a long-term RV survey
of nearby  low-mass stars using several spectrometers.   Most data are
obtained  at  the  1.65-m  telescope  at the  Moletai  observatory  in
Lithuania  \citep{S16}.   A  CORAVEL-type  spectrometer  was  used  to
measure the  RVs with a  typical accuracy of  the order of \kms  and a
spectral resolution around 20000.  HIP~43431 was observed with CORAVEL
at Moletai  several times in the period  from 2000 to 2014.  Owing to the
relative faintess of the star and the complex multi-line nature of its
spectrum, the CCF dips are noisy  and often blended. In this paper, we
do not use these CORAVEL observations.

In  2015,   a  modern fibre-fed  echelle  spectrometer   VUES  \citep{VUES}  was
commissioned  at  Moletai.   We  took  spectra  of  HIP~43431  with  a
resolution of  30\,000 in  the wavelength range  from 400  to 880\,nm.
The spectra recorded by the  CCD detector are extracted and calibrated
in   the  standard   way.    The  RV   is   determined  by   numerical
cross-correlation of  the spectrum with  a binary mask,  emulating the
CORAVEL method in software (Fig.~\ref{fig:VUES}).  Compared to CORAVEL, the RVs delivered by
VUES  are more  accurate; their  rms  residuals from  the orbits  are,
typically,  from 0.2  to  0.3  \kms.  For  each  observing night,  the
instrumental  velocity  zero  point  and  its drift  were  checked  by
observations  of a  few RV  standard stars.   The mean  RV zero-point,
calculated  using   186  measurements   of  the  standard   stars,  is
$\Delta$RV=0.09$\pm$0.01  \kms,  and the  standard  deviation is  0.19
\kms.  One  can suspect a small  drift of the zero-point  from 0.04 to
0.14  \kms  in about  three  years.   Practically  the same  value  of
$\Delta$RV=0.08$\pm$0.05  \kms  (rms 0.18  \kms,  $n=15$) is  obtained
using telluric lines in the spectra of HIP 41431 as the RV reference.

\subsubsection{CHIRON observations}
\label{subsubsec:CHIRON}

\begin{figure}
\centerline{
\includegraphics[width=8.5cm]{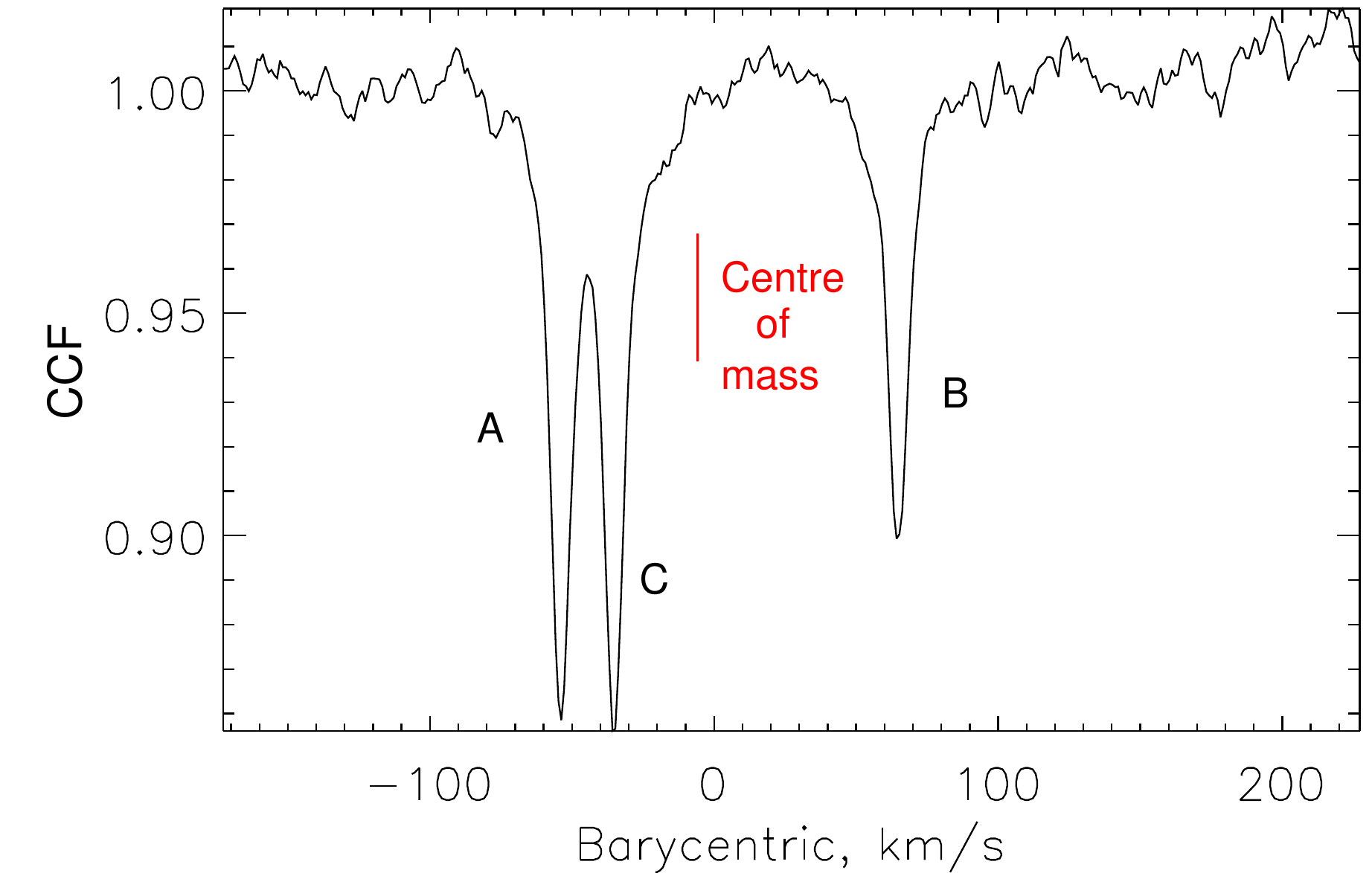}
}
\caption{CCF  of HIP  41431  using  the CHIRON  spectrum  taken on  JD
  2458443.8. 
\label{fig:CHIRON} }
\end{figure}

Seven spectra of HIP~41431 were taken at the 1.5-m telescope located at Cerro Tololo
(Chile)  and operated  by  the SMARTS  consortium.\footnote{See 
\url{http://www.astro.yale.edu/smarts/}}  Observations  were
conducted by the telescope operator  in the service mode.  The optical
echelle spectrometer CHIRON \citep{CHIRON} was used in the slicer mode
with  a  spectral  resolution  of  80000.  On  each  visit,  a  single
10-minute exposure of the star  was taken, accompanied by the spectrum
of  the comparison  lamp for  wavelength calibration.   The  data were
reduced by the pipeline written in IDL.

The RVs are derived from the reduced spectra by cross-correlation with
a binary  mask based on the  solar spectrum, similarly  to the CORAVEL
RVs.  More details are  provided by \citet{Tok2016}. Only the spectral
range from 4500\AA ~to 6500\AA, relatively free from telluric lines, was used for
the CCF calculation.  The RVs delivered by this procedure should be on
the  absolute  scale  if   the  wavelength  calibration  is  good.   A
comparison of  CHIRON RVs with  several RV standards revealed  a small
offset  of +0.16  km~s$^{-1}$ \citep{Tok2018};  in the  following this
offset   is   neglected.   

Figure~\ref{fig:CHIRON} illustrates  the 3-component cross-correlation
function  (CCF) derived from  the CHIRON  spectrum. The  strongest dip
belongs to the component  C; the dip of A is almost  equal, while B is
obviously weaker. The relative dip areas of C:A:B are 1:0.95:0.71. The
dips are narrow and  correspond to the projected rotational velocities
of  5.5,   5.1,  and  4.2   \kms  according  to  the   calibration  of
\citet{Tok2016}.  The rotation of  A and  B is almost two times slower than
synchronous (10.1 and 9.9 \kms for the primary and the secondary, respectively.  The first four sets of RVs measured with CHIRON are plotted in Fig.~\ref{fig:orb} together with the spectro-photodynamical model curves (see Sect.\,\ref{sec:dyn}).   

\subsubsection{UVES archival spectra}
\label{sec:UVES}

In  an  effort to  extend  the time  coverage,  we  consulted the  ESO
archive\footnote{\url{http://archive.eso.org/cms.html}}   and   found  eight
high-resolution spectra  taken with UVES  at the 8-m VLT  telescope in
December  2017,  on  two  nights,  in the  framework  of  the  program
0100.D-0282(A)   to   study   chromospheric   activity   of   inactive
main-sequence stars  (PI A. Santerne). The  data recently
became  public. We  measured the  RVs using  only the  red-arm spectra
(wavelength  range  5655--9463\AA)  by  correlation  with  the  binary
mask. No zero-point correction  was applied. These RVs serve primarily
to confirm the  4-year modulation of the centre-of-mass  RV induced by
the star D.

\subsection{Gaia astrometry}

The  {\it Gaia} data  release 2,  DR2 \citep{Gaia},  provides accurate
parallax  (see  Table~\ref{tab:object})  and  proper  motion  (PM)  of
HIP~41431. However, the  reduced goodness-of-fit parameter {\tt gofAL}
of 27.19  and the statistically  significant excess noise of  0.29 mas
show that the  single-star model adopted in DR2  is not adequate.  The
photo-centre position is modulated  with the 59-day and 4-year periods
of  the middle and  outer orbits,  and the  future data  releases will
hopefully provide the astrometric elements of these orbits.

Comparison of  the DR2 position  with the second {\it  Hipparcos} data
reduction  \citep{HIP2}  allows  us  to  compute  the  average  PM  of
$(\mu_\alpha, \mu_\delta)_{\rm mean}= (+10.62, +21.95)$ mas~yr$^{-1}$.
This long-term PM  agrees well with the ground-based  PM of $(+9.2 \pm
1.1, +21.1  \pm 1.1)$ mas~yr$^{-1}$ given in  UCAC4 \citep{UCAC4}, but
differs very  significantly from the ``instantaneous''  PM measured by
{\it Gaia}:  $\Delta \mu_{\rm DR2-mean}  = (-7.00 \pm 0.20,  -1.12 \pm
0.14)$   mas~yr$^{-1}$.   A   similar,   although  less   significant,
difference  $\Delta \mu  _{\rm HIP2-mean}=  (-10.7 \pm  3.3,  -1.6 \pm
2.5)$ mas~yr$^{-1}$ is found between the {\it Hipparcos} and long-term
PM.  So, HIP~41431 is an  astrometric binary of the $\Delta \mu$ type.
We  show  below  that  the  measured  $\Delta  \mu$  is  explained  by
the photocentric motion induced by the outer orbit.

\subsection{Speckle interferometry}

The star was observed in 2018.97  in the $I$ band using speckle camera
at the  4.1-m Southern  Astrophysical Research (SOAR)  telescope.  The
angular  resolution (minimum detectable  separation) was  50\,mas, and
the dynamic  range (maximum  magnitude difference) was  about 4  mag at
0\farcs15  separation.   The instrument  and  observing technique  are
described in \citet{SOAR}. No companions  were detected. The star D is
too faint  compared to  the combined light  of ABC and,  moreover, its
estimated  separation  at  the  moment  of the  observation  was  only
30\,mas. Nevertheless, speckle interferometry is still useful to probe
the absence of additional resolved companions. However,
  \citet{ohetal17}   found   two    co-moving   stars,   HIP~37165   and
  TYC~2468-87-1, both at  projected separations of  $\sim$9\,pc.  Given the
  separation,  they  cannot  be  bound companions  of  HIP~41431.  The
  spatial motion (Table~\ref{tab:object}) is typical for the Galactic disk population.  

\section{Orbits and masses}
\label{sec:orb}

\begin{table*}
\center
\caption{Provisional orbital elements of the inner triple}
\label{tab:orb}
\medskip
\begin{tabular}{l ccc }
\hline
Element & CfA & VUES   & CHIRON  \\
\hline
$P_1$ (d)           & 2.93326$\pm$0.0003  & 2.93291    *          &    2.93291 *         \\
$\tau_1$ (BJD $-$2400000) & 51833.92$\pm$0.26   & 58213.890$\pm$0.003  & 58478.701$\pm$0.096   \\
$e_1$               & 0.019$\pm$0.005     & 0.011 *               &  0.015 $\pm$0.003   \\
$\omega_1$ (deg)    & 23.5$\pm$45.8       & 77.5  *               &  182.0$\pm$11.7 \\
$K_1$ (\kms)        & 78.54$\pm$0.55      & 79.55$\pm$0.39       & 79.73$\pm$0.69  \\
$K_2$ (\kms)       & 80.45$\pm$0.69       & 80.4$\pm$0.41        & 80.36$\pm$0.69 \\
$M_{\rm A,B} \sin^3 i_{\rm A,B}$ (${\cal M}_\odot$) & 0.618, 0.603 &  0.631, 0.621 & 0.652, 0.620  \\ 
$P_2$  (d)         & 58.819$\pm$0.037     & 58.963 *               &  58.963 *  \\
$\tau_2$ (BJD $-$2400000)& 51820.20$\pm$1.61    & 58186.997$\pm$0.066  & 58481.39$\pm$0.44  \\
$e_2$             & 0.275$\pm$0.007       & 0.2749 *              & 0.2749 *  \\
$\omega_2$ (deg) &  112.9$\pm$3.1         & 177.4 *               & 178.3$\pm$2.4  \\
$K_3$  (\kms)    &  25.23$\pm$0.66        & 23.84$\pm$0.30       & 23.72$\pm$0.31  \\
$K_4$ (\kms)     &  49.79$\pm$1.02       & 46.98$\pm$0.21        & 46.86$\pm$0.46   \\
$\gamma$ (\kms)  &  $-$6.37$\pm$0.22     & $-$12.23$\pm$0.14     & $-$8.34$\pm$0.14  \\
$\sigma_{\rm A,B,C}$ (\kms)& 2.03, 2.60, 1.66 & 0.46, 1.20, 0.31 & 0.14, 0.17, 0.27 \\
$M_{\rm AB,C} \sin^3 i_{\rm AB,C}$ (${\cal M}_\odot$) & 1.517, 0.769 &  1.278, 0.648  & 1.266,0.641   \\
\hline
\end{tabular}
\end{table*}

\begin{figure}
\centerline{
\includegraphics[width=8.5cm]{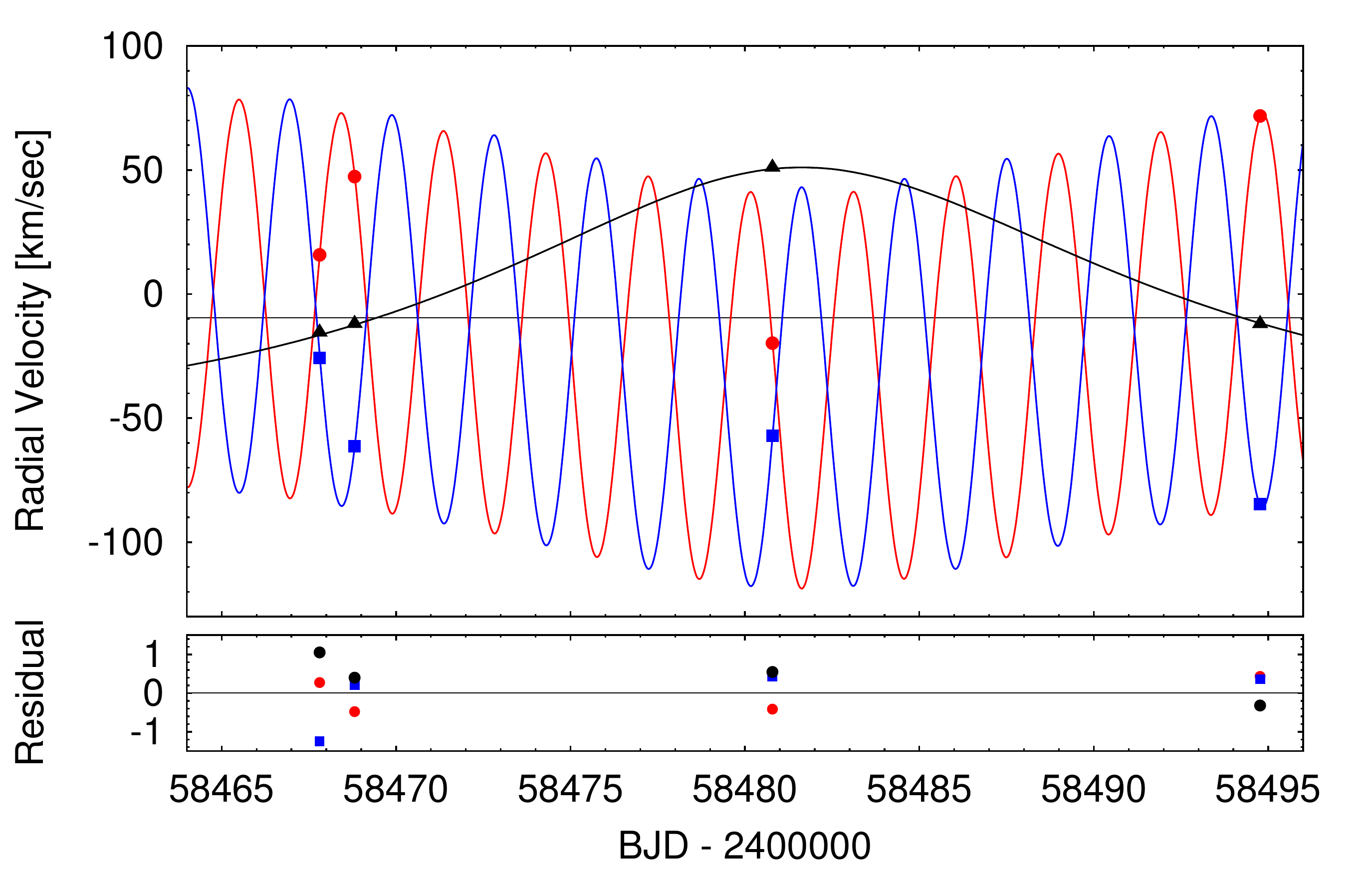}
}
\caption{ A recent, one-month-long section of the RV curves of all the three 
spectroscopically visible components. Red circles, blue squares and black triangles
represent the observed data of stars A, B and C, respectively, while the lines
with similar colors show the full spectro-photodynamical model solution (see Sect.\,\ref{sec:dyn}).
In the bottom panel the residual values are plotted. The largest residuals correspond to spectra with blended lines.
\label{fig:orb} }
\end{figure}

Dynamical interaction  between the inner  and outer orbits  means that
the observed RVs cannot be accurately modelled as superposition of two
Keplerian  orbits.  Dynamical  modelling  using both  RVs  and ETV  is
presented  in  the  following Sect.\,\ref{sec:dyn}.  However,  fitting
Keplerian orbits to the subsets of RVs provided important insights and
led  to the  discovery of  the fourth  star,  D.  Table~\ref{tab:orb}
lists spectroscopic  orbital elements of the inner  and middle systems
derived  from three  independent  sets of  RVs  coming from  different
instruments.  Both  orbits were  fitted simultaneously using  the {\tt
  orbit3.pro} IDL code \citep{TL2017}.  Some elements were fixed (they
are listed with asterisks instead of errors). The RV amplitudes of the inner pair A,B are denoted as
$K_1$ and $K_2$, the RV amplitudes in the outer orbit are $K_3$ (center of mass of AB) and $K_4$. 

The orbits in the first column of Table~\ref{tab:orb} were computed by
D.   L. in 2008  based on 30  RVs measured  with the  CfA from  2007.2 to
2008.3,  using his  own code  for fitting  two  orbits simultaneously.
Incomplete  phase coverage  of  the outer  orbit  likely explains  the
slight  disagreement  of  the   RV  amplitudes  $K_3$  and  $K_4$  and
corresponding masses with  the recent orbits based on  VUES and CHIRON
data. For the latter, we  fixed the periods and the outer eccentricity
to  their values determined  photometrically. 
The  VUES RVs measured before 2018.2 were corrected for the offsets due 
to the outer orbit (see below).

The  most striking  disagreement  between those  orbits concerns  the
centre-of-mass  velocity   $\gamma$.  The  velocity   zero  points  of
respective spectrographs are carefully controlled, hence the effect is
real. The centre-of-mass velocity $V_0$ can be computed for each
individual observation independently of the orbital elements as
\begin{equation}
V_0 = (M_A V_A + M_B V_B + M_C V_C)/(M_A+ M_B + M_C)
\label{eq:v0}
\end{equation}
using relative component's masses  derived from the RV amplitudes.  We
adopted provisionally $M_A:M_B:M_C  = 1:0.98:1.007$  and applied eq.~\ref{eq:v0}  to the
observations where all three RVs  are measured from the  same spectrum,
excluding  spectra  with  blended  lines.   The  correct  choice  of
the relative  masses is  verified by  the absence  of  correlation between
$V_0$ and RVs of the individual components.

\begin{figure}
\centerline{
\includegraphics[width=8.5cm]{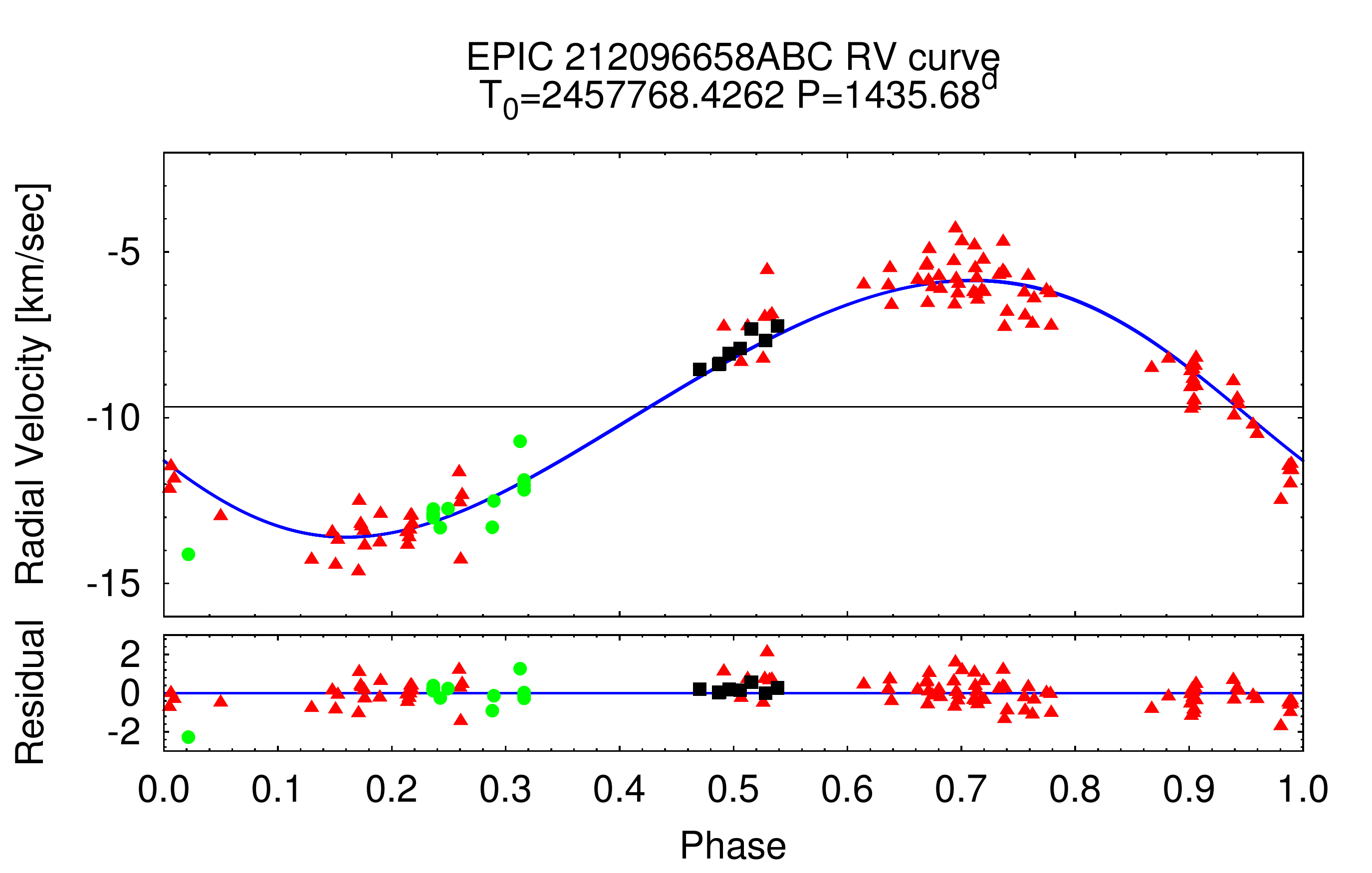}
}
\caption{ RV curve of the centre-of-mass ABC corresponding to the outer
  orbit with $P_3 = 3.9$ yr.  Red triangles plot the CfA RVs, black squares
  -- CHIRON RVs, green circles -- other RVs. Blue line represents the RV curve
calculated from the spectro-photodynamical solution. The residuals to this solution are shown in the bottom panel. 
\label{fig:out} }
\end{figure}

A plot of $V_0$ vs. time  clearly shows its variation with a period of
$\sim$4 years; three cycles between 1999.3 and 2019.0 are covered. The
preliminary  orbital elements  describing the  $V_0(t)$  variation are
given   in   Table~\ref{tab:out},  the   RV   curve   is  plotted   in
Fig.~\ref{fig:out}. The RV amplitude in the outer orbit is denoted by $K_5$. 
We adopted the erors of 1 \kms  for the CfA data,
0.5 \kms for TRES, VUES and  UVES, and 0.1 \kms for CHIRON. The latter
RVs  are distinguished  by the  tight sequence  of  black squares  around the
phase  0.5, showing the  $V_0$ trend  in just  two months.  The global
weighted rms residuals are 0.38  \kms. The long and extensive coverage
of the CfA data is essential for constraining the outer orbit.

\begin{table}
\center
\caption{Elements of the outer orbit}
\label{tab:out}
\medskip
\begin{tabular}{l c}
\hline
Element & Value \\
\hline
$P_3$ (d)           & 1427$\pm$5 \\
$\tau_3$ (BJD $-$2400000)          & 51786$\pm$102 \\  
$e_3$               & 0.068$\pm$0.022 \\
$\omega_3$ (deg)    & 52$\pm$27 \\
$K_5$ (\kms)        & 3.963$\pm$0.15 \\
$\gamma$ (\kms)     & $-$9.82$\pm$0.10 \\
\hline
\end{tabular}
\end{table}

The middle and  inner orbits computed from the  CfA data of 2007--2008
happen to be  near the maximum of the  RV curve in Fig.~\ref{fig:out},
hence $\gamma_{\rm  CfA} =  -6.4$ \kms;  the  trend during  this period  was
small.  Similarly,  most VUES observations  cover the minimum  of this
curve,  hence  $\gamma_{\rm  VUES}  =  -12.2$  \kms  (the  first  VUES
observation does not match the inner orbits without an offset correction). 

Given that the inner pair is eclipsing, the factor $\sin^3 i_{\rm A,B}
\approx 1$, so  the spectroscopic masses are close  to the true masses
of the  components.  The mass  sum of 1.9  \msun for A+B+C and  the RV
amplitude of the  outer orbit then lead to $M_D >  0.36$ \msun. If the
inclination  of  the outer  orbit  were  substantially different  from
$90^\circ$,   the   large  resulting   $M_D$   would  contradict   its
non-detection in the  spectra, so we adopt $M_D =  0.4$ \msun (this is
confirmed below by the full modelling).  The total mass and the period
define the semimajor axis of the outer orbit, 3.26 au or 67 mas on the
sky.

At the {\it Gaia} DR2 epoch, 2015.5, the star D was receding from ABC,
moving  toward maximum  separation. The  projected speed  of  the mean
orbital  motion  during the  time  interval  of  2015.5$\pm$0.5 years  was
$\mu_{\rm orb}  = 40$ mas~yr$^{-1}$,  directed away from  the primary.
Comparing this speed to the  observed $\Delta \mu = 7.0$ mas~yr$^{-1}$
and neglecting the light of the  star D allows a direct measurement of
the outer mass  ratio $q_3$ from the relation $  \Delta \mu / \mu_{\rm
  orb} = q_3/(1 + q_3)$.  Hence,  $q_3 = 0.21$ and $M_D = 0.40$ \msun.
This  confirms  our  assumption  that  the outer  orbit  has  a  large
inclination.   The  direction  of   $\Delta  \mu$  suggests  that  the
companion D  was at the position  angle of $\sim  80^\circ$ in 2015.5.
Without  actually   resolving  the  outer  binary,   we  already  know
approximately all its orbital elements and can compute the positions.

\section{Dynamical modelling}
\label{sec:dyn}

As a consequence of the compactness of this quadruple system (the period ratios are $P_2/P_1\sim20.2$ and $P_3/P_2\sim24.4$) the orbital motions of the four stars depart significantly from pure Keplerian orbits. Therefore, the accurate modelling of all  observations  needs a spectro-photodynamical approach, i.\,e. the combination of the simultaneous analysis of the RVs and photometric data with the numerical integration of the four-body motion. This analysis was carried out with the software package {\sc Lightcurvefactory} \citep[see][and further references therein]{Borkovits2019} of which the latest version is now able to handle quadruple systems both in 2+2 and 2+1+1 configurations. The relevant modifications of the orbital equations to be numerically integrated in this new version are discussed in Appendix\,\ref{app:numint}. 

Apart from the inclusion of the fourth star forming the third, outermost
``binary'' with the centre of mass of the ABC components, this complex
analysis was carried out in a very similar manner as  described in
Sect.\,7 of \citet{Borkovits2019} and, therefore, here we discuss only
the basic steps briefly. We carried out a joint  Markov Chain Monte
Carlo (MCMC)  parameters search for the following data series:
\begin{itemize}
\item[(i)] Two sets of long cadence {\em K2} lightcurves;
\item[(ii)] The RVs of components A, B, and C;
\item[(iii)] The ETV curves of the innermost EB (for both primary and secondary minima).
\end{itemize}

Regarding item (i), we consider two variants of processed {\em K2} lightcurves, with constant and variable eclipse depth, as described in Sect.\,\ref{subsubsec:K2obs}. In both cases, we  use only a narrow window of width $\sim0.12$\,d centered on each eclipse (blue points in Fig.~\ref{fig:eclipsefit}, right panel). The out-of-eclipse brightness variations are negligible, and the omission of these data saves a significant amount of computational time. Most dynamical  information coded in the lightcurves is contained in the fine structure and timings of the eclipses. Note also, that for the $\sim29.4$-min long-cadence time of {\em Kepler}, we apply a cadence time correction on the model lightcurves \citep[see][for details]{Borkovits2019}. 
Considering that the eclipse depth variation in the  {\em Kepler} data has been confirmed by our  ground-based photometry,  we discuss below  only the uneven eclipse depth solution.

Turning to the RV curves, we emphasize that instead of  fitting the
usual analytical formulae, our numerical integrator calculates for all
time instances the 3D velocity vectors of all four bodies, the $v_z$
components of those vectors give directly the RVs relative to the
centre of mass of the quadruple system. The systemic velocity,
$\gamma$, is then calculated a posteriori by a simple linear
regression minimizing the $\chi^2_\mathrm{RV}$  residuals between
measured RVs and the model. We found only minor zero-point differences
amongst the RV instruments (see Sect.\,\ref{subsec:spectroscopy}) and
neglected  them. 

In principle, the  {\em K2} lightcurves carry the same timing information as the ETV curves, making the latter redundant. However, the advantages of using both the lightcurves and the ETV curves together have been explained in \citet{Borkovits2019}. Similarly to our previous work, the ETV curves were used to preset the period ($P_1$) and phase term $(\mathcal{T}_0)_1$ of the innermost binary for each new set of the trial parameters. The latest ETV points  from the ground-based photometry were also added to the data set.

During our analysis, we carried out several dozens of MCMC runs and
tried different sets and combinations of adjustable parameters. We
also applied some additional relations to constrain some of the
parameters in order to reduce the degrees of freedom in our
problem. For example, while the masses of all  four stars  can be
deduced from the joint dynamical analysis of the ETV and RV curves
and, combining these results with the outputs of the lightcurve
analysis, the physical dimensions of the eclipsing stars can also be
determined, none of the observational data used for the photodynamical
modelling carry information on the radii of the stars C and
D. Similarly, only the temperature ratio of stars B and A
($T_\mathrm{B}/T_\mathrm{A}$) can be constrained by the light curve,
while the effective  temperature of one star in the inner binary
should be taken from an external source. The photodynamical model can
say nothing on the effective  temperatures of the stars C and D. Their
net flux in the {\em Kepler} band is manifested only as extra flux for
the lightcurve model. In order to get reliable information on these
parameters we applied different options. Regarding $T_\mathrm{A}$, in
some runs we constrained it with a Gaussian prior centered to  the
temperature given in {\em Gaia} DR2 ($T_\mathrm{eff}=3978$\,K), while
in another series of runs the code calculated internally the
temperature in each trial step from the stellar mass $m_\mathrm{A}$
with the use of the mass--temperature relations of \citet{Tout96},
valid for zero age main sequence (ZAMS) stars. During our analysis the radii of the two outer stars ($R_\mathrm{C,D}$) and also the effective temperature of the star D ($T_\mathrm{D}$) were also connected internally to their masses via the relations of \citet{Tout96}. (For these calculations solar metallicity was assumed.) Applying these three constraints, the fourth remaining parameter, i.~e. $T_\mathrm{C}$, takes the role of the extra light parameter ($l_\mathrm{x}$) and, therefore, there is no need to use this latter one.

Besides the above mentioned constraints, in most of our runs we adjusted the following parameters: 
\begin{itemize}
\item[(i)]{Three parameters related to the orbital elements of the
    inner binary: eccentricity ($e_1$), the phase of the secondary
    eclipse relative to the primary one ($\phi_\mathrm{sec,1}$) which
    constrains the argument of periastron ($\omega_1$, see
    \citealt{Rappaport2017}), and the inclination ($i_1$)\footnote{ Note, again, that $P_1$ and $(\mathcal{T}_0)_1$ are constrained through the ETV curves.};}
\item[(ii)-(iii)]{Two times six parameters related to the orbital
    elements of the middle and the outermost orbits: $P_{2,3}$,
    $(e\sin\omega)_{2,3}$, $(e\cos\omega)_{2,3}$, $i_{2,3}$, the times
    of the periastron passages of stars C and D along their
    revolutions on the middle and the outermost orbits, respectively
    ($\tau_{2,3}$), and the position angles of the nodes of the two orbits ($\Omega_{2,3}$)\footnote{Strictly speaking, as we set $\Omega_1=0\degr$ at epoch $t_0$ for all runs, adjusting the other two $\Omega_{2,3}$-s is practically equivalent to the adjustment of the differences of the nodes (i.~e., $\Delta\Omega$-s), which are the truly relevant parameters for dynamical modelling.};}
\item[(iii)]{Four mass-related parameters: the mass of the component A, $m_\mathrm{A}$, and the mass ratios of all  three orbits $q_{1,2,3}$;}
\item[(iv)]{and, finally, four other parameters which are related
    (almost) exclusively to the lightcurve solutions, as follows: the
    duration of the primary eclipse $(\Delta t)_\mathrm{pri}$ closest
    to epoch $t_0$ (which is an observable that is strongly connected
    to the sum of the fractional, i.~e. scaled by the inner semi-major axis,
    radii of stars A and B, see \citealt{Rappaport2017}), the ratio of the radii of stars A and B ($R_\mathrm{B}/R_\mathrm{A}$), and the temperature ratios of $T_\mathrm{B}/T_\mathrm{A}$ and $T_\mathrm{C}/T_\mathrm{A}$.} 
\end{itemize}

 Turning to other, lightcurve-dependent parameters, we applied a logarithmic limb-darkening law, where the coefficients were interpolated from the pre-computed passband-dependent tables in the {\sc Phoebe} software \citep{Phoebe}. The {\sc Phoebe}-based tables, in turn, were derived from the stellar atmospheric models of \citet{castellikurucz04}. Due to the nearly spherical stellar shapes in the inner binary, an accurate setting of gravity darkening coefficients has no influence on the lightcurve solution and, therefore, we simply adopted a fixed value of $g=0.32$ which is appropriate for late-type stars according to the traditional model of \citet{lucy67}.  We also found that the illumination/reradiation effect was quite negligible for the eclipsing binary; therefore, in order to save computing time, this effect was neglected. On the other hand, the Doppler-boosting effect \citep{loebgaudi03,vankerkwijketal10} was included into our model.
 Furthermore, in the absense of any other information, we assumed that
 the equatorial planes of stars A and B are aligned with the innermost
 orbital plane. The projected rotational velocities of stars A, B and
 C were set to their spectroscopically obtained values (see
 Sect.\,\ref{subsubsec:CHIRON}).\footnote{ These settings are
   irrelevant for the lightcurve modelling of such almost spherical
   stars, but matter for  the longer-term dynamical studies discussed in Sect.\,\ref{sec:orbprop}.} 

\begin{figure*}
\begin{center}
\includegraphics[width=0.85\textwidth]{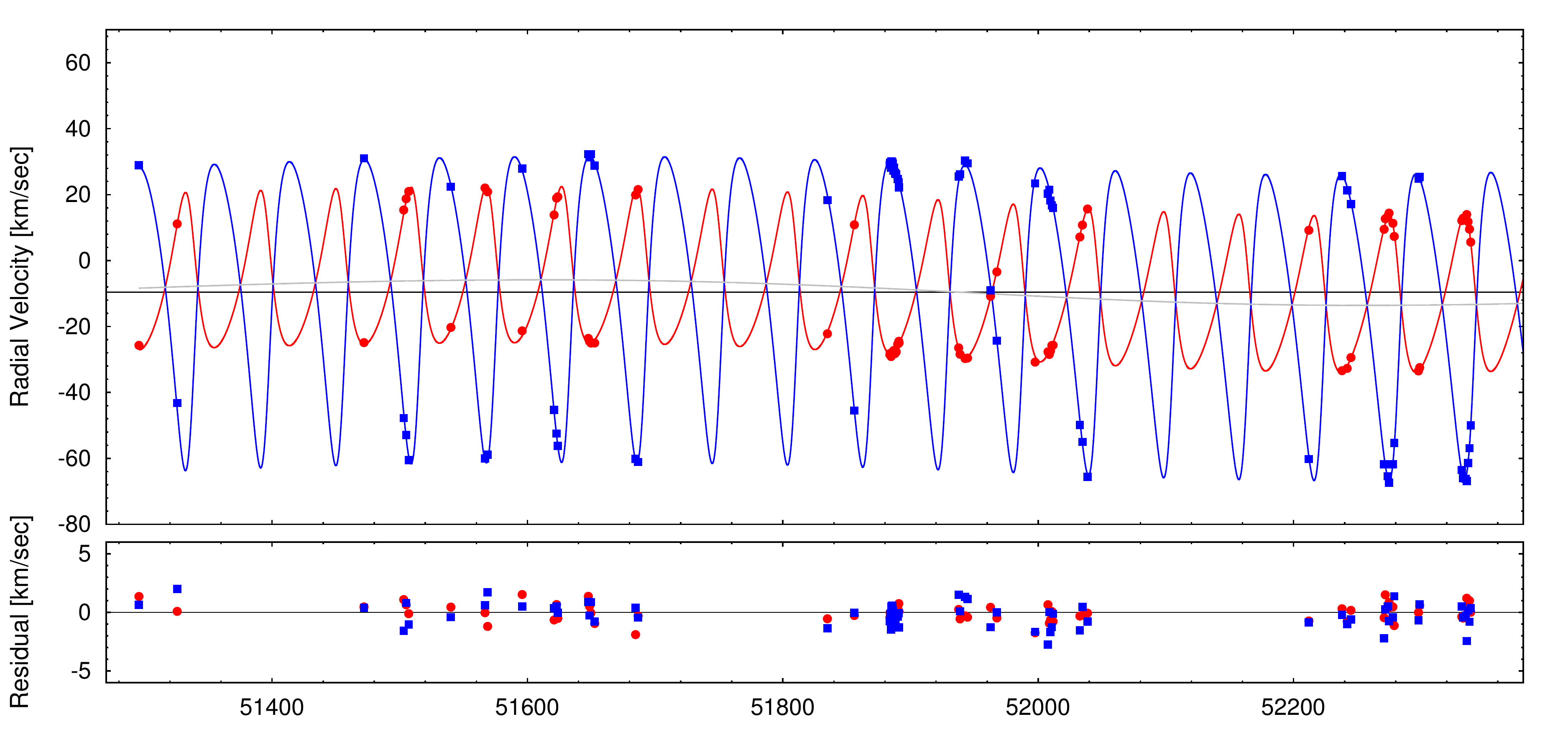}
\includegraphics[width=0.85\textwidth]{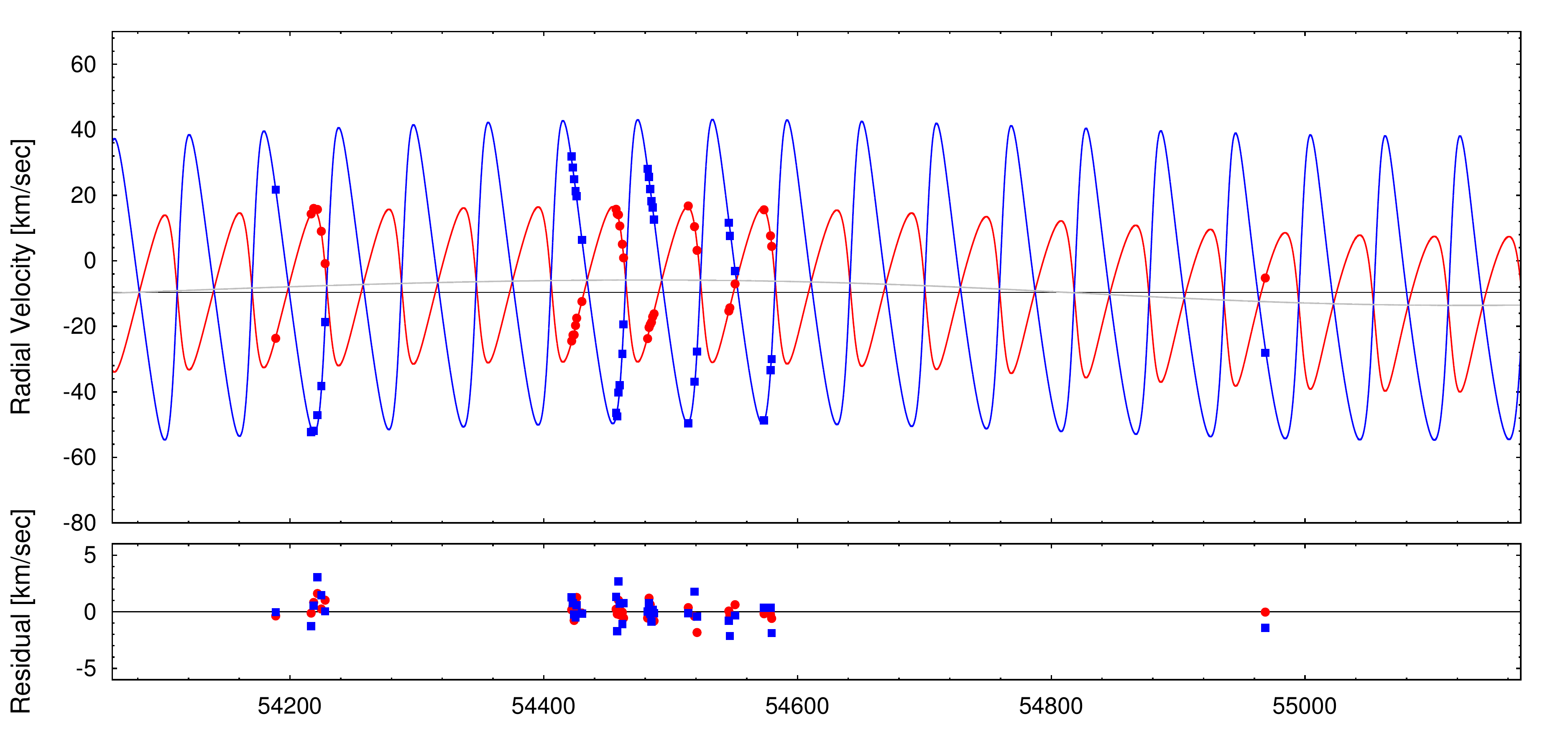}
\includegraphics[width=0.85\textwidth]{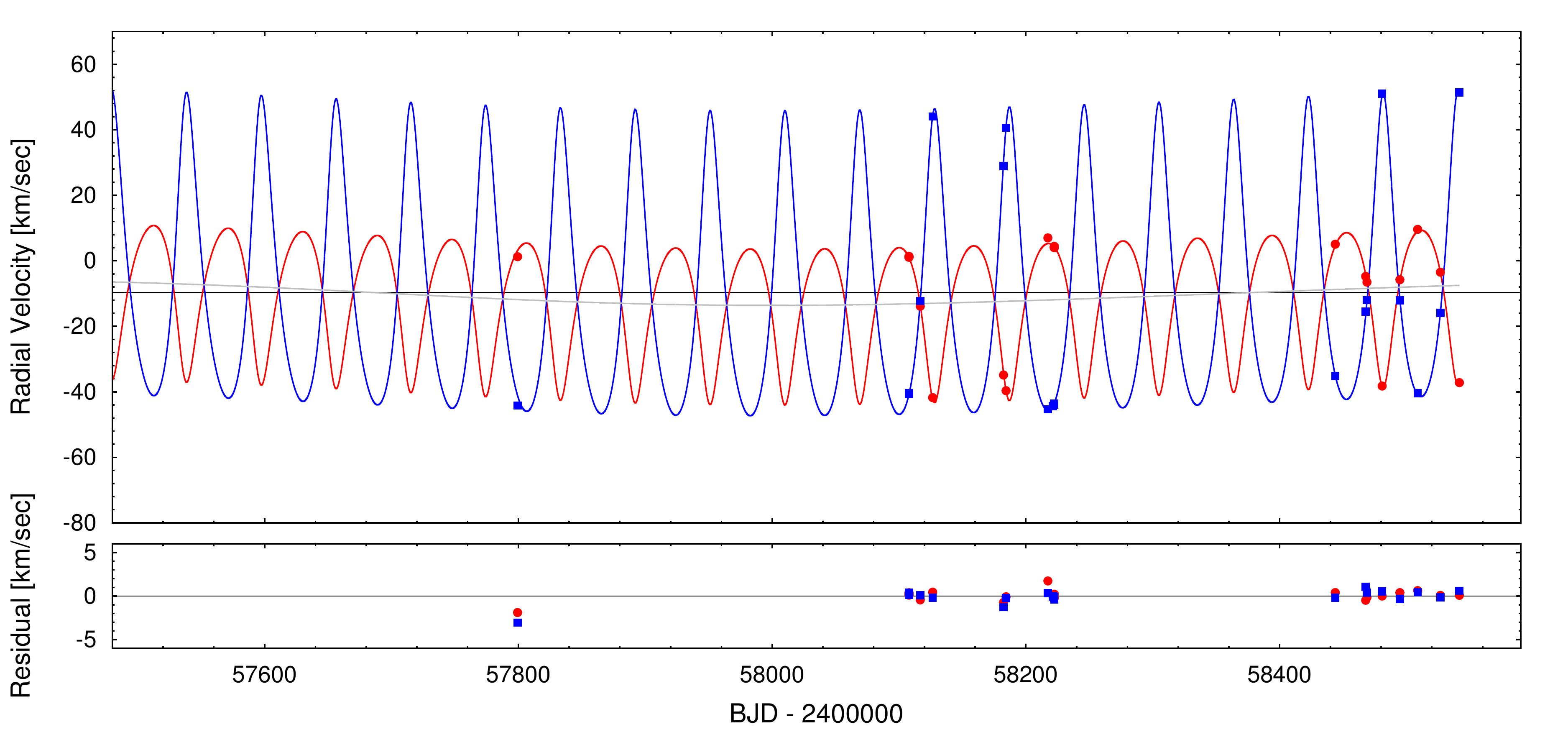}
\caption{ 1100-day-long sections of RV data, model
  solution, and residuals. For better visibility, we do not plot the
  RVs of the innermost pair.  Red circles denote
  the mass-weighted average of the observed RVs of stars A and B
  (i. e. the ``observed'' RV of their barycentre), while blue boxes
  stand for the directly observed RV data of star C. Red and blue
  lines show the appropriate model RV curves. Furthermore, grey line
  represent the outermost orbit RV component, while thin black line
  stands for the constant $\gamma$ velocity. Note 
  variations both in the shape and orientation of the RV orbits. These are consequences of the quick apsidal motion of the middle orbit due to the strong dynamical interactions of the four stars.  }
\label{fig:RVmiddleouter} 
\end{center}
\end{figure*}

The orbital and astrophysical parameters derived from the `uneven eclipse depth scenario' spectro-photodynamical analysis are tabulated in Table\,\ref{tab: syntheticfit}, and will be discussed in the subsequent Sections\,\ref{sec:par} and \ref{sec:orbprop}. The corresponding model lightcurves are presented in Fig.\,\ref{fig:eclipsefit}, while the different RV curves  are shown in Figs.\,\ref{fig:orb}, \ref{fig:out}, and \ref{fig:RVmiddleouter}. Finally, the model ETV curve plotted against the observed ETVs is shown in Fig.\,\ref{fig:ETV}.

\begin{table*}
 \centering
 \caption{Orbital and astrophysical parameters from the joint photodynamical lightcurve, three RV curves and ETV solution}
 \label{tab: syntheticfit}
\begin{tabular}{@{}lllll}
  \hline
\multicolumn{5}{c}{orbital elements$^a$} \\
\hline
   & \multicolumn{4}{c}{subsystem} \\
   & \multicolumn{2}{c}{A--B} & AB--C & ABC--D \\
  \hline
  $P$ [days] & \multicolumn{2}{c}{$2.93001\pm0.00008$} & $59.157\pm0.006$ & $1441.4\pm10.7$  \\
  $a$ [R$_\odot$] & \multicolumn{2}{c}{$9.259\pm0.054$} & $78.77\pm0.48$ & $700.9\pm6.7$ \\
  $e$ & \multicolumn{2}{c}{$0.00873\pm0.00005$} & $0.2781\pm0.0040$ & $0.0814\pm0.0034$ \\
  $\omega$ [deg]& \multicolumn{2}{c}{$175.96\pm6.32$} & $332.87\pm0.51$ & $293.04\pm2.87$ \\ 
  $i$ [deg] & \multicolumn{2}{c}{$87.725\pm0.032$} & $86.528\pm0.136$ & $84.170\pm5.073$\\
  $\tau$ [BJD - 2400000]&\multicolumn{2}{c}{$57142.7234\pm0.0515$} & $57123.285\pm0.089$ & $57768.4\pm10.1$ \\
  $\Omega$ [deg] & \multicolumn{2}{c}{$0.0$} & $-1.797\pm0.088$ & $-21.422\pm16.598$ \\
  $i_\mathrm{m}$ [deg] & \multicolumn{2}{c}{$-$} & $2.157\pm0.107$ & $21.658\pm16.345$\\
  \hline
  mass ratio $[q=m_\mathrm{sec}/m_\mathrm{pri}]$ & \multicolumn{2}{c}{$0.983\pm0.009$} & $0.511\pm0.009$ & $0.186\pm0.019$ \\
  $K_\mathrm{pri}$ [km\,s$^{-1}$] & \multicolumn{2}{c}{$79.217\pm0.593$} & $23.687\pm0.318$ & $3.861\pm0.333$ \\ 
  $K_\mathrm{sec}$ [km\,s$^{-1}$] & \multicolumn{2}{c}{$80.599\pm0.599$} & $46.353\pm0.404$ & $20.706\pm0.428$ \\ 
  $\gamma$ [km\,s$^{-1}$] & \multicolumn{3}{c}{$-$} & $-9.608\pm0.129$ \\
  \hline  
\multicolumn{5}{c}{stellar parameters} \\
\hline
   & A & B &  C & D\\
  \hline
 \multicolumn{5}{c}{Relative quantities} \\
  \hline
 fractional radius [$R/a$]  & $0.0635\pm0.0012$ & $0.0622\pm0.0012$  & $0.0075\pm0.0002 $ & $0.00048\pm0.00004$\\
 fractional flux [in {\em Kepler}-band]& $0.3232$  & $0.2867$    & $0.3550$ & $0.0351$ \\
 \hline
 \multicolumn{5}{c}{Physical Quantities} \\
  \hline 
 $m$ [M$_\odot$] & $0.625\pm0.010$ & $0.614\pm0.012$ & $0.633\pm0.016$ & $0.349\pm0.036$\\
 $R$ [R$_\odot$] & $0.588\pm0.012$ & $0.576\pm0.012$ & $0.590\pm0.015^b$ & $0.336\pm0.025^b$\\
 $T_\mathrm{eff}$ [K]& $4043\pm60^c$ & $3986\pm60$ & $4064\pm83$ & $3373\pm159^c$\\
 $L_\mathrm{bol}$ [L$_\odot$] & $0.0829\pm0.0060$ & $0.0753\pm0.0055$ & $0.0852\pm0.0078$ & $0.0131\pm0.0032$\\
 $M_\mathrm{bol}$ & $7.44\pm0.08$ & $7.55\pm0.08$ & $7.41\pm0.10$ & $9.44\pm0.26$\\
 $M_V           $ & $8.50\pm0.11$ & $8.69\pm0.12$ & $8.44\pm0.14$ & $12.24\pm0.73$\\
 $\log g$ [dex] & $4.70\pm0.02$ & $4.71\pm0.02$ & $4.70\pm0.03$ & $4.93\pm0.08$ \\
 \hline
$(M_V)_\mathrm{tot}$             &\multicolumn{4}{c}{$7.34\pm0.07$} \\
distance$^d$ [pc]                &\multicolumn{4}{c}{$<50.2\pm1.7$}  \\  
\hline
\end{tabular}

{\em Notes. }{$a$: Instantaneous, osculating orbital elements, calculated for epoch $t_0=2457143.395$ (BJD); $b$: Calculated from the mass--radius relations of \citet{Tout96}; $c$: Calculated from the mass--temperature relations of \citet{Tout96}; $d$: Photometric maximum distance, see text for details.}
\end{table*}

As a sanity check for the photodynamical solution, we calculate the
maximum photometric distance of HIP\,41431 by combining the total
$V$ magnitude of the system  (see the penultimate row
in Table\,\ref{tab: syntheticfit}) with the apparent $V$ magnitude (see Table\,\ref{tab:object}).
This  results in a photometric distance of $d_\mathrm{phot}\leq50\pm2$\,pc, which
is in good agreement with the {\em Gaia} parallax.

\section{Physical parameters of the components}
\label{sec:par}

\begin{figure}
\centerline{
\includegraphics[width=8.5cm]{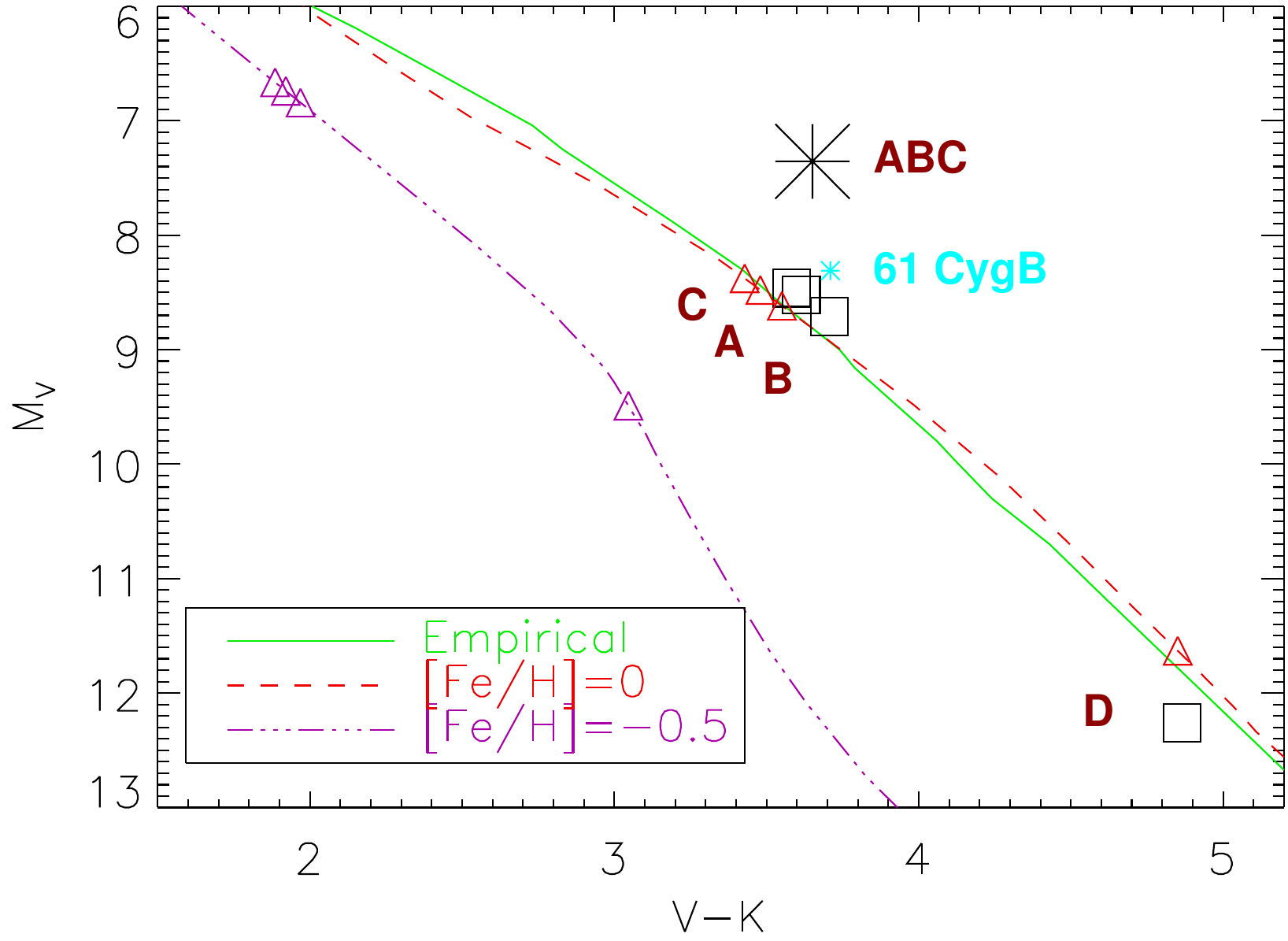}
}
\caption{Color-magnitude diagram. The  lines are 1-Gyr isochrones from
  \citet{PARSEC} for metallicity [Fe/H] of 0 and $-0.5$ dex and the empirical relations of \citet{Pecaut2013}. 
Squares show  the  components, the  triangles are
  isochrone locations  corresponding to the measured  masses, the large
  asterisk is the combined light  of all stars. The location of the nearby K7V dwarf 61 Cyg B is shown by the blue star.
\label{fig:cmd} }
\end{figure}

The masses obtained from the spectro-photodynamical model are 
in good agreement with masses deduced from the preliminary RV analysis
and the {\em Gaia} astrometry (Sect.\,\ref{sec:orb}).
Similarly, the relative fluxes of the visible stars A, B, and C in the
$V$ band deduced from the model, 0.95:0.79:1, 
match well their relative fluxes measured spectroscopically (see Sect.~\ref{sec:CfA}). The minor contribution of the faint star D to the $V$-band flux 
is accounted for by assuming that it is an MS dwarf.  The $V-K$ colors of the stars
are computed from their effective temperatures listed in Table~\ref{tab: syntheticfit} using standard relations 
for MS stars \citep[e.g.][]{Pecaut2013} and adjusted to match the measured combined $V-K_s$ color in Table~\ref{tab:object}. 
This allows us to place the stars on the color-magnitude diagram (CMD) in Fig.~\ref{fig:cmd} using the {\em Gaia} parallax.

The masses, colors, and  absolute magnitudes of
the visible stars A, B, and  C match well both the empirical relations of \citet{Pecaut2013}
and the theoretical isochrone for  solar metallicity, while the star D
of  0.35 \msun   contributes only 0.01 to the  total light in the
$V$ band and 0.35 in the {\em Kepler} band. The empirical relations 
of \citet{Benedict2016} for M-type dwarfs predict the $V-K$ colors from 3.61 to 3.66 mag for stars with masses of the components A, B, and C, and match the observed combined color $V - K_s = 3.65$ mag. The observed absolute $V$ magnitudes are brighter than those of Benedict et al. by $\sim$0.3 mag, either because these stars are slightly evolved or because of the reduced blanketing owing to sub-solar metallicity.  The
stars A, B, C have  effective temperature close to 4000\,K or slightly lower ({\it Gaia}
gives $T_{\rm eff}=3978$\,K) and gravity  $\log g = 4.7$ in cgs units.
The PARSEC isochrone for [Fe/H]=$-0.5$ \citep{Palacios2010},  on the  other hand,  corresponds to
bluer and  brighter stars  (for the same  masses) and  contradicts the
observations.  The discrepancy between theoretical isochrones and actual colors of low-mass stars has been recently 
noted by \citet{Howes2019}. They discuss  the ``benchmark'' K7V star 61~Cyg~B (HD~201092) as an example. 
Its parameters and  position on the CMD happen to be similar to the components A, B, and C of HIP~41431. 
The measurements of [Fe/H] for the 61 Cyg A and B listed in Simbad have a large scatter, illustrating the difficulty 
of the spectroscopic analysis of late-type dwarfs. Most measurements give  [Fe/H]$\approx -0.4$ dex for 61 Cyg, 
and the metallicity of HIP~41431 is likely similar, i.~e. mildly sub-solar.

\begin{figure}
\centerline{
\includegraphics[width=8.5cm]{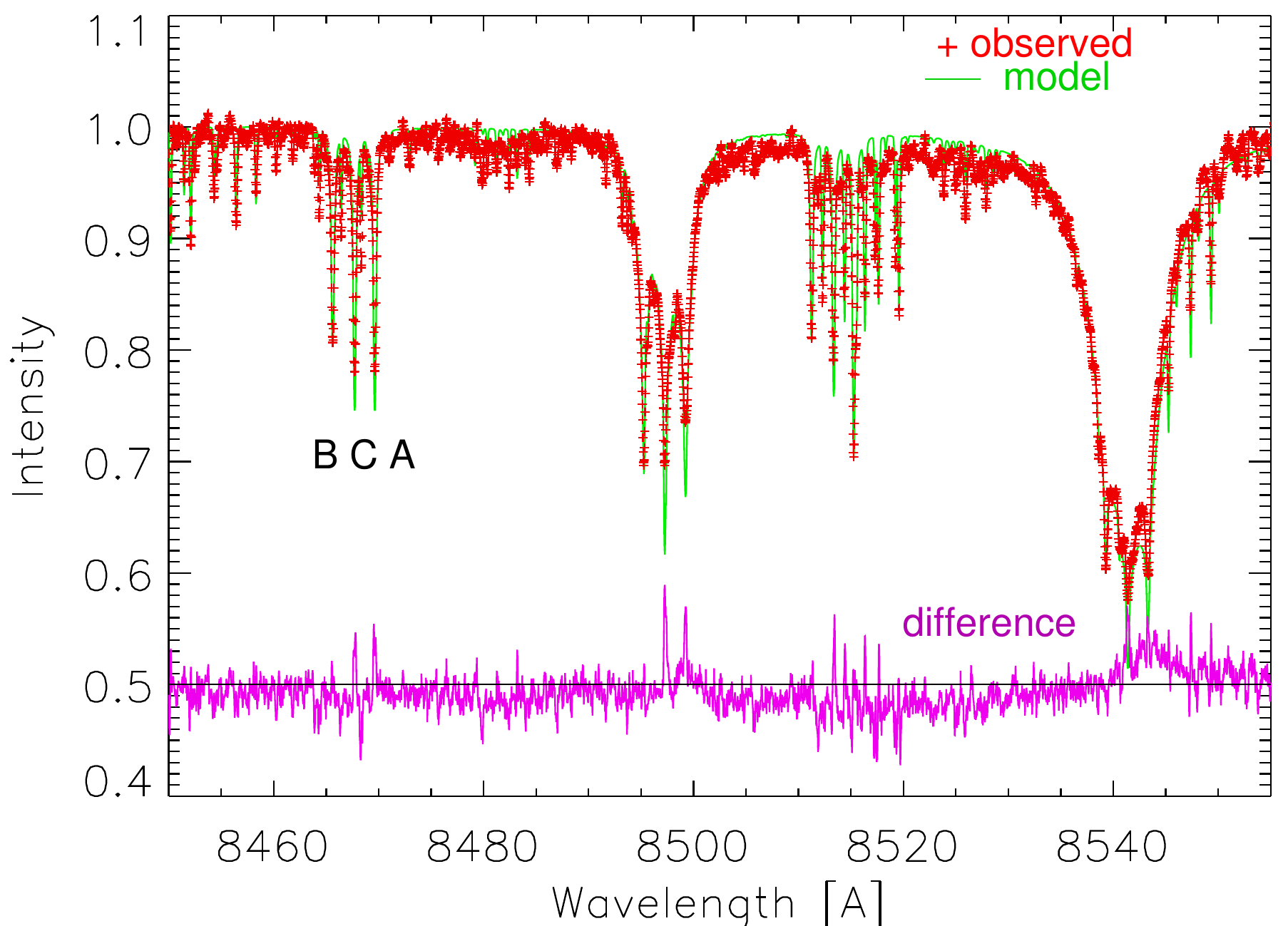}
}
\caption{Observed spectrum (red crosses)  is compared to the synthetic
  spectrum   with  $T_{\rm   eff}=4000$\,K,  $\log   g  =   4.7$,  and
  [Fe/H]=$-$0.5  dex  (green line)  
  around the CaII near-infrared triplet. The
  magenta line  shows the difference.  The components' order  (left to
  right) is B, C, A.
\label{fig:UVES} }
\end{figure}

Stellar parameters  can be measured directly  from the high-resolution
spectra  if  the spectra  of  individual  components  are  isolated
(disentangled). This approach was  implemented for the CHIRON spectra,
but even by combining them  all and averaging the component's spectra,
the signal to noise ratio (SNR) remains modest.  Alternatively, we can
compute  the triple-lined  synthetic spectrum  and compare  it  to the
observed  one.  The  spectrum of  HIP~41431 with  the largest  SNR was
taken   on  2017  December   29  (JD   2458116.697)  with   UVES  (see
Sect.~\ref{sec:UVES}). Using the measured  RVs and relative fluxes, we
shift  and scale  the  synthetic spectrum  to  model the  triple-lined
system.   Such forward-modelling  avoids the  need to  disentangle the
observed spectrum.  A small correction  for the estimated  dilution by
the light of the star D is applied.

Figure~\ref{fig:UVES} shows a fragment of  the near-IR UVES
spectrum  compared   to  the   synthetic  spectrum  from   the  Pollux
library\footnote{See \url{http://pollux.oreme.org}} \citep{Palacios2010}. We
made this comparison for the [Fe/H] values ranging from $-1$ to $+0.5$ dex
and found  the best match for  [Fe/H]=$-$0.5 dex. Note that the effective temperature of the synthetic spectrum 
 differs slightly from  the estimated stellar temperatures, but
 synthetic spectra of dwarfs with lower temperatures are not available
 in the Pollux library. 

We  tried to  find  the spectral  signature  of the  faint  star D  by
correlating  the   difference  between  the  UVES   spectrum  and  its
triple-lined  model with  the  synthetic spectra  of different  $T_{\rm
  eff}$ or  with a  binary mask, near  the CaII infrared  triplet. The
resulting   correlation  function  does   not  contain   the  expected
details. The  star D could  have a fast  rotation or could be  a close
spectroscopic pair.  

\section{Orbital properties and dynamical evolution}
\label{sec:orbprop}

As mentioned previously,  the 
four stars in this compact quadruple system interact dynamically and therefore, the three orbits 
are subjected to strong and fast dynamical perturbations. Spectacular manifestations of these interactions are nicely visible even 
on the four-year-long data train of the measured ETVs (Fig.\,\ref{fig:ETV}) of the innermost, eclipsing pair. 

First, we note the 59-day sine-like modulation of the ETV, similar (but not identical) for the primary and secondary ETVs.  The dominating contributor to this modulation is the third-body perturbation from the star C that alternates the mean motion (and also the orbital elements) of the innermost binary on the timescale of the period of the middle orbit \citep{Borkovits2015}. The contribution of the classic light-travel time effect (LITE) to this 59-day variation is only about 10\%. 

\begin{figure*}
\begin{center}
\includegraphics[width=0.49 \textwidth]{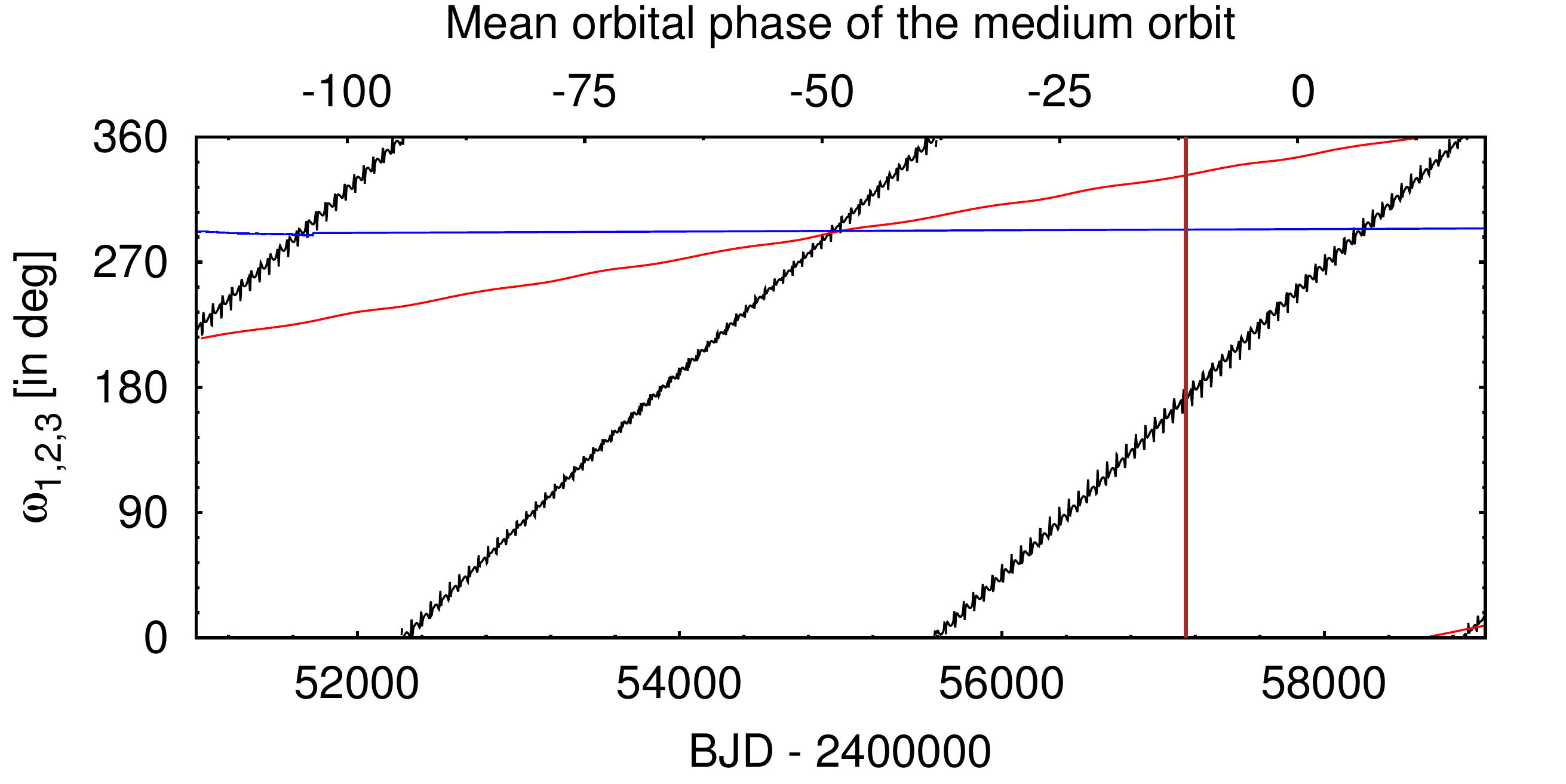}
\includegraphics[width=0.49 \textwidth]{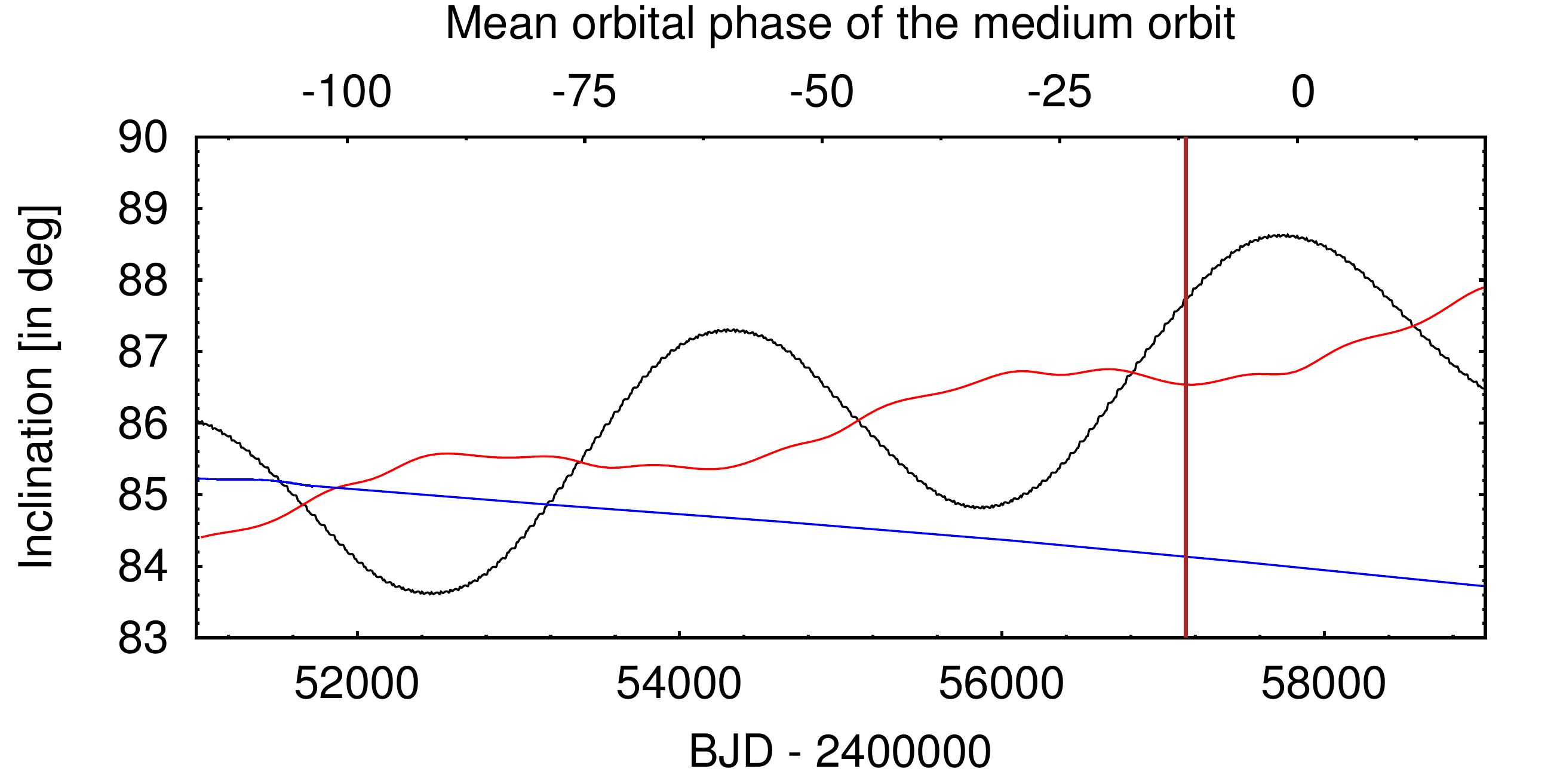}
\caption{Variations of the osculating (observable) arguments of periastron ($\omega_{1,2,3}$) and inclinations ($i_{1,2,3}$) of the three orbits between 1998.5 and 2020.4, as  calculated from our spectro-photodynamical solution. Black, red and blue lines denote the orbital elements of the innermost, middle and outer orbits, respectively. The elements are averaged for the period of the corresponding orbits. Vertical brown line represents the beginning of the {\em Kepler} observations.}
\label{fig:orbelements_numint} 
\end{center}
\end{figure*}

Second, the crossing of the two ETV curves reveals a fast, dynamically forced apsidal motion. According to our numerical integration, which was a substantial part of the photodynamical solution, during this four years the major axis of the innermost orbit has turned by $\approx150-160\degr$, i.\,e., has made almost half a revolution (see Fig.\,\ref{fig:orbelements_numint}). Therefore, the current period of the apsidal motion of the innermost orbit is $U_1\approx9$\,yr. Only two binaries formed by non-degenerate stars with  shorter apsidal motion periods are known to date. These are the inner binaries of the compact hierarchical triple systems KOI-126 and  KIC 05771589, reported by \citet{carteretal11} and \citet{Borkovits2015}, respectively.

Finally, the $P_3\sim3.9$\,yr  LITE  induced by the outermost component is also present. 

 At this point we have to note, that the ETV residuals (Fig\,\ref{fig:ETV}, lower panel) show small, but systhematic departures in the order of some $10^{-4}$\,days around BJD 2\,458\,150 and 2\,458\,520, i.~e. during the late campaign (C16 and C18) observations of the {\em K2} mission. In this moment we cannot decide whether these small, but systhematic discrepancies, which however do not exceed the estimated accuracies of the individual ETV points have physical origins indicating some inaccuracies in the parameters of the expected four-body model, or they are consequences of some instrumental effects. 

Turning to the RV data, a more spectacular, and almost uniquely observed manifestation of the apsidal rotation of the middle orbit can be seen in Fig.\,\ref{fig:RVmiddleouter}, where we plot the RVs of  the three visible stars A, B, and C (after subtracting the orbit of the eclising pair),
together with the corresponding spectro-photodynamical model for the whole, 20-year-long time span of our observations. The apsidal rotation of the middle orbit results in the notable variation of the shape  of the RV curves.\footnote{ While spectroscopically detected apsidal motions were previously reported for other close binaries \citep[see e.~g.][]{ferreroetal13} and even for an exoplanet \citep{csizmadiaetal19}, too, we are not aware of any other systems where such a significant fraction of a complete apsidal revolution period was covered with RV data so densely as in the present case.}

\begin{figure}
\begin{center}
\includegraphics[width=0.49 \textwidth]{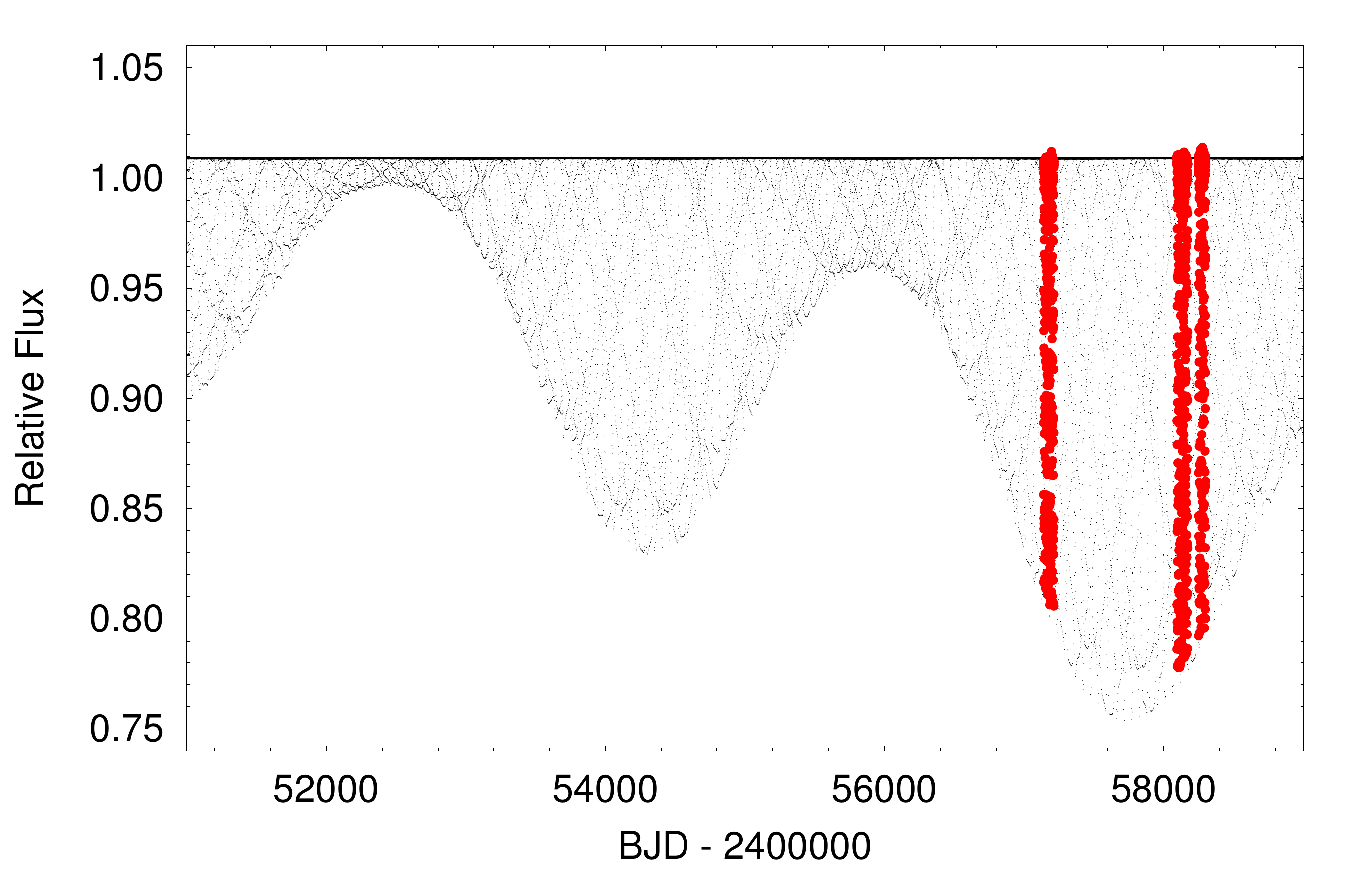}
\caption{ The model lightcurve of the star between 1998.5 and 2020.4 (thin black curve). The eclipse depth variations caused by 
the orbital plane precession are well visible. The {\em K2} lightcurves collected during Campaigns 5, 16 and 18 are also plotted with red circles.}
\label{fig:lclong} 
\end{center}
\end{figure}

Orbital inclination is another key observable in an eclipsing binary. Our photodynamical solution has revealed small, 
but definitely non-zero relative (mutual) inclination between the two inner orbital planes:   $i_\mathrm{mut1-2}=2\fdg2\pm0\fdg1$.\footnote{Note, that the combination of the dynamical and geometrical effects on the light- and ETV curves break the degeneracy between prograde and retrograde solutions, therefore an almost coplanar, but retrograde solution can be ruled out with high confidence.}
The mutual inclination between the outermost and the  inner and middle orbits is larger, although its uncertaintly is substantial:  
 $i_\mathrm{mut1-3}=22\degr\pm16\degr$ and $i_\mathrm{mut2-3}=20\degr\pm16\degr$.  
The non-coplanarity of the orbits triggers precession of all three orbital planes,  illustrated in the right panel of Fig.\,\ref{fig:orbelements_numint}, where the variations of the three observable orbital inclinations (i.\,e. the angles between the orbital planes and the plane of the sky) are plotted.

While the dynamical effects of such orbital misalignments are expected to occur only on very long times scales, their observational consequences, however, are manifested almost promptly, in dramatic eclipse depth variations. 
In Fig.\,\ref{fig:lclong} we plot the model lightcurve of the system
since the beginning of the spectroscopic observations. The
$\approx9.2$\,year period cyclic variation of the eclipse depths
which, naturally, correlates with the $\Delta i_1\approx3.5\degr$
amplitude, short-term variation of the inclination ($i_1$) of the
innermost orbit, is clearly visible.\footnote{The precession period is
  in perfect agreement with the analytically calculated period  within
  the framework of the stellar three-body problem \citep[see,
  e.\,g.,][Eq.\,27]{soderhjelm75}. This fact illustrates that on short
  time scales, the effects of the third and fourth bodies remain
  almost independent, at least, from a dynamical point of view.}
Moreover, another  (on this timescale linear) effect is also well
visible; it corresponds to the longer time-scale and larger-amplitude
precession triggered by the more inclined outermost orbit. As a
consequence, if the photodynamical solution is correct, in the
forthcoming decades one can expect that the mean visible inclinations
of the innermost and middle orbits (i.\,e., $i_1$ and $i_2$ averaged
over the $\approx9.2$\,yr-period of the short-term precession) and,
therefore, the averaged eclipse depths will increase. Moreover, when
these inclinations reach $\approx90\degr$ around 2040,  eclipses of
the component C should also become observable for several years.

We  emphasize, however, that the relative nodal angle of the outermost
orbit ($\Omega_3$) is obtained only with a large uncertainty and thus,
the corresponding two outer mutual inclinations are only weakly
constrained. Therefore, these results should be considered as
tentative. The reason of this  uncertainty is that only the eclipse
depth variation of the eclipsing pair is strongly sensitive to the
rate of the (visible) inclination variation of the innermost pair and,
therefore, only the {\em K2} observations, which cover a small
fraction of the $\approx1260$\,day-long interval contain really
conclusive information about the several hundred-year-long outer
precession cycle. Moreover, as  discussed in
Sect.\,\ref{subsubsec:K2obs},  the reality of the
eclipse depth variations observed by  {\em Kepler} might be debatable. Therefore, follow-up observations and continuous monitoring of the eclipse depth variations are crucial.

\begin{figure*}
\begin{center}
\includegraphics[width=0.49 \textwidth]{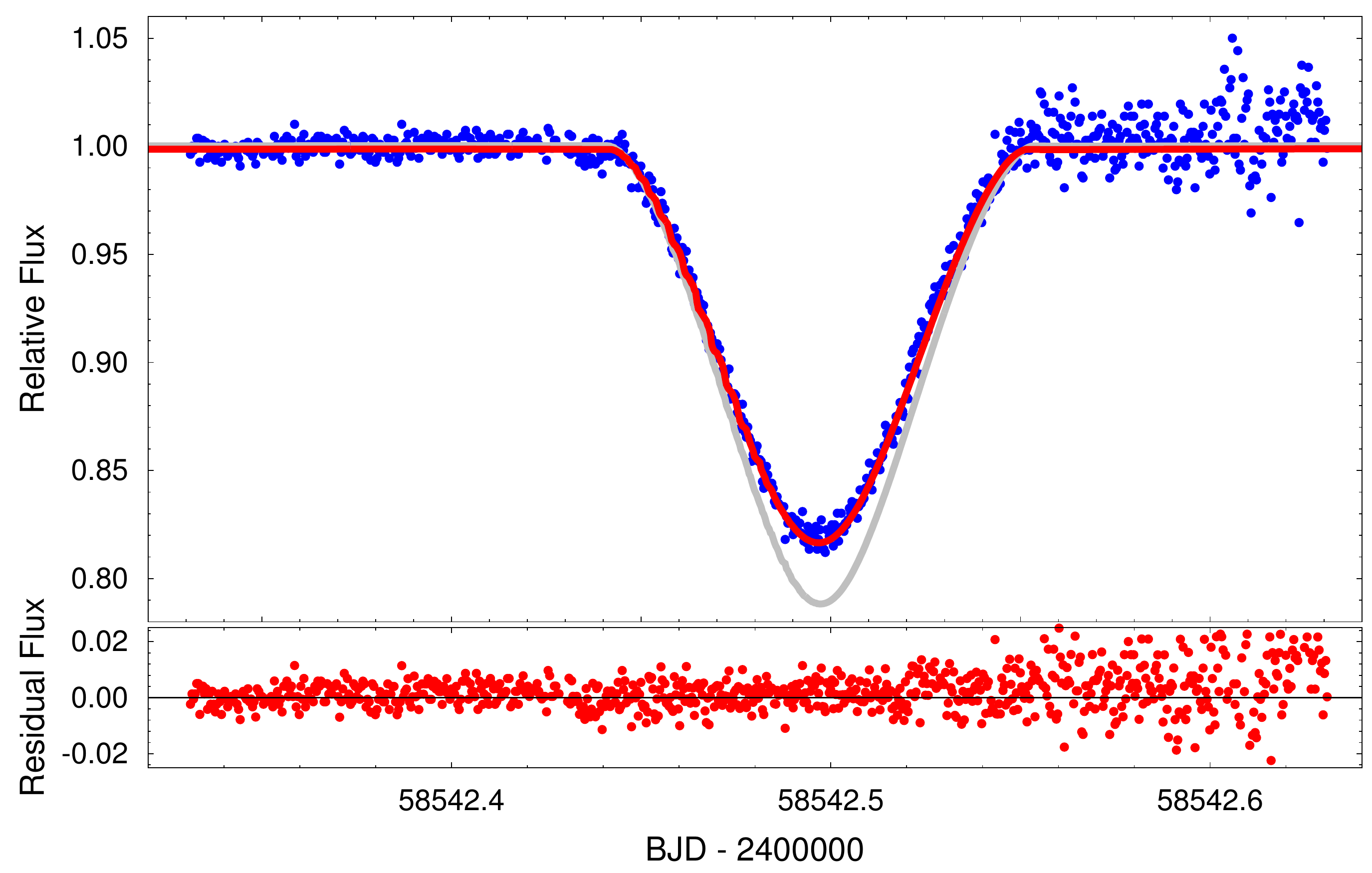}\includegraphics[width=0.49 \textwidth]{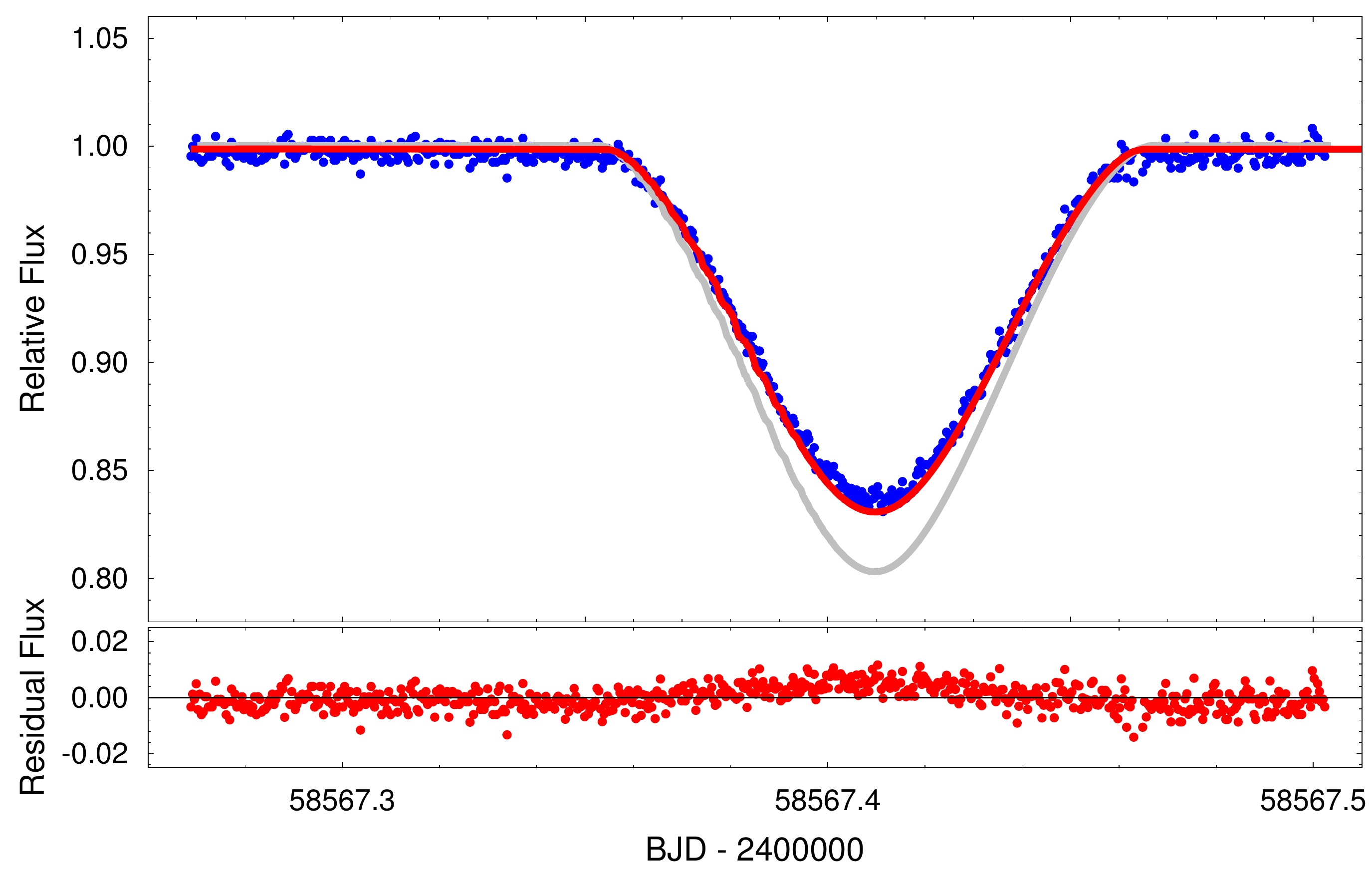}
\caption{ {\em Upper left panel:} Primary eclipse of HIP\,41431 measured in $R_C$-band on the night of 27/28 Feb 2019 (blue circles) and the corresponding photodynamical model lightcurves both for the uneven and uniform {\em K2} eclipse depths scenarios (red, and grey lines, respectively). As one can see, the measured eclipse depth is in perfect agreement with the predictions of the uneven eclipse depths scenario and therefore, it confirms the physical origin of the eclipse depth differences amongst the different {\em K2} campaigns.  {\em Upper right panel:} The same for the secondary eclipse measured also in $R_C$-band on the night of 24/25 March 2019. {\em Lower panels:} Observations vs. uneven eclipse depth model lightcurve residuals.}
\label{fig:lcRc} 
\end{center}
\end{figure*}

We added this star to the long-term eclipse monitoring programme of
Baja Observatory, Hungary, as a top priority target. Unfortunately,
due to the bad weather conditions (which are usual in the winter
season), so far  we were able to observe only four primary  and two
secondary eclipses. Furthermore, owing to the poor sky conditions, two
 primary minima were observed in unfiltered mode, and  only four eclipses were observed with a standardized Kron-Cousins $R_C$
filter. Normally, unfiltered minima observations are useful for the
times of minima determination but unfit for studying the eclipse depth
variation. Therefore, we consider only the $R_C$-band eclipse
observations. We generated the $R_C$-band model lightcurve for those
nights and compared it to the observations (see
Fig.\,\ref{fig:lcRc}). As one can see, the agreement for the primary
eclipse is almost perfect. For the secondary eclipse, a minor
systematic deviation can be seen. However, the decrease of the  eclipse
depths is beyond doubt. This fact confirms not only the ongoing precession of the innermost orbit but, retrospectively, justifies the physical origin of the eclipse depth variations observed in the different {\em K2} campaigns.  

 In order to check the long time-scale dynamical evolution and stability of our quadruple system, we carried out further numerical integration on a timescale of $10^8$\,yr. The integrator was the same as in the case of the photodynamical analysis. Therefore, beyond the four-body point-mass forces, tidal forces acting upon in the innermost binary were also considered,  including the Eulerian equations of the rotations of stars A and B.  Furthermore, for some additional runs tidal dissipation (within the framework of the equilibrium tide approximation), and relativistic apsidal motion were also included (see Appendix\,\ref{app:numint} for details). 
The additional parameters necessary for these integrations were set as follows. The inner structure (or apsidal motion) constants of both stars A and B were set to $k_2=0.02$ which, according to \citet{torresetal10},  is appropriate for such low-mass stars. Furthermore, the dissipation rates for both stars were set to $\lambda=2\times10^{-5}$. This type of dissipation rate was defined by Eq.\,(13) in \citet{kiselevaetal98}. It is connected to the small tidal lag time through the formula:
\begin{equation}
\Delta t=-\frac{3}{8}\sqrt{\frac{R^3}{Gm}}\left(1+2k_2\right)^2\lambda
\end{equation}
(see \citealp{borkovitsetal04}, Eq.\,25). The choosen numerical values of $\lambda$ correspond to tidal lags of $\Delta t\approx-2\times10^{-10}$\,days for both stars.
The integrations did not reveal any dramatic variations in the orbital elements of the three orbits. Therefore, we conclude that the orbital configuration is stable up to the nuclear evolution time scales.

On the other hand, the numerical method allows us to study the spin evolution of the innermost two stars. This is especially interesting in the present case, as the most unusual  characteristic of  this  system is  the slow  axial rotation of the stars comprising  the inner pair.\footnote{The  Referee, however, noted that Fig.\,6  of \citet{Lurie2017} contains other eclipsing binaries with short periods and substantially sub-synchronous rotation measured from starspots.}  For HIP\,41431, the standard assumption that the axes are perpendicular to the orbit is not trivial. The likely non-coplanarity of the outermost orbit forces significant orbital plane precession, which may lead to spin-orbit misalignement, as  suggested e.~g. by \citet{beustetal97} in the case of TY CrA. Furthermore, as found by \citet{correiaetal16}, the secular evolution of the spins in hierarchical triple systems when viscous tidal forces are present might be affected strongly by secular resonances between orbital and spin precessions and, therefore, chaotic rotation might  occur. 

 In what follows we discuss briefly some results of three different
 integrator runs. Dissipative forces were taken into account in all
  three runs. For the run `A', the spin axes of stars A and B are
 parallel to the orbital spin vector of the innermost orbit at the
 epoch $t_0$ used in the photodynamical model. Furthermore, the spin
 rates are set according to the spectroscopically measured projected
 rotation velocities. In other words, apart from the dissipation terms,
 this numerical integration is a simple extention of the accepted
 photodynamical model over a much longer time scale. For the run `B', the
 initial orbital elements were the same, but the orientation and
 magnitude of stellar spins are set arbitrarily. Finally, for the run `C',
 the initial parameters are the same as in run `A', but the outermost,
 fourth body is removed, i.~e., a three-body integration was carried out.
 
\begin{figure*}
\begin{center}
\includegraphics[width=0.49 \textwidth]{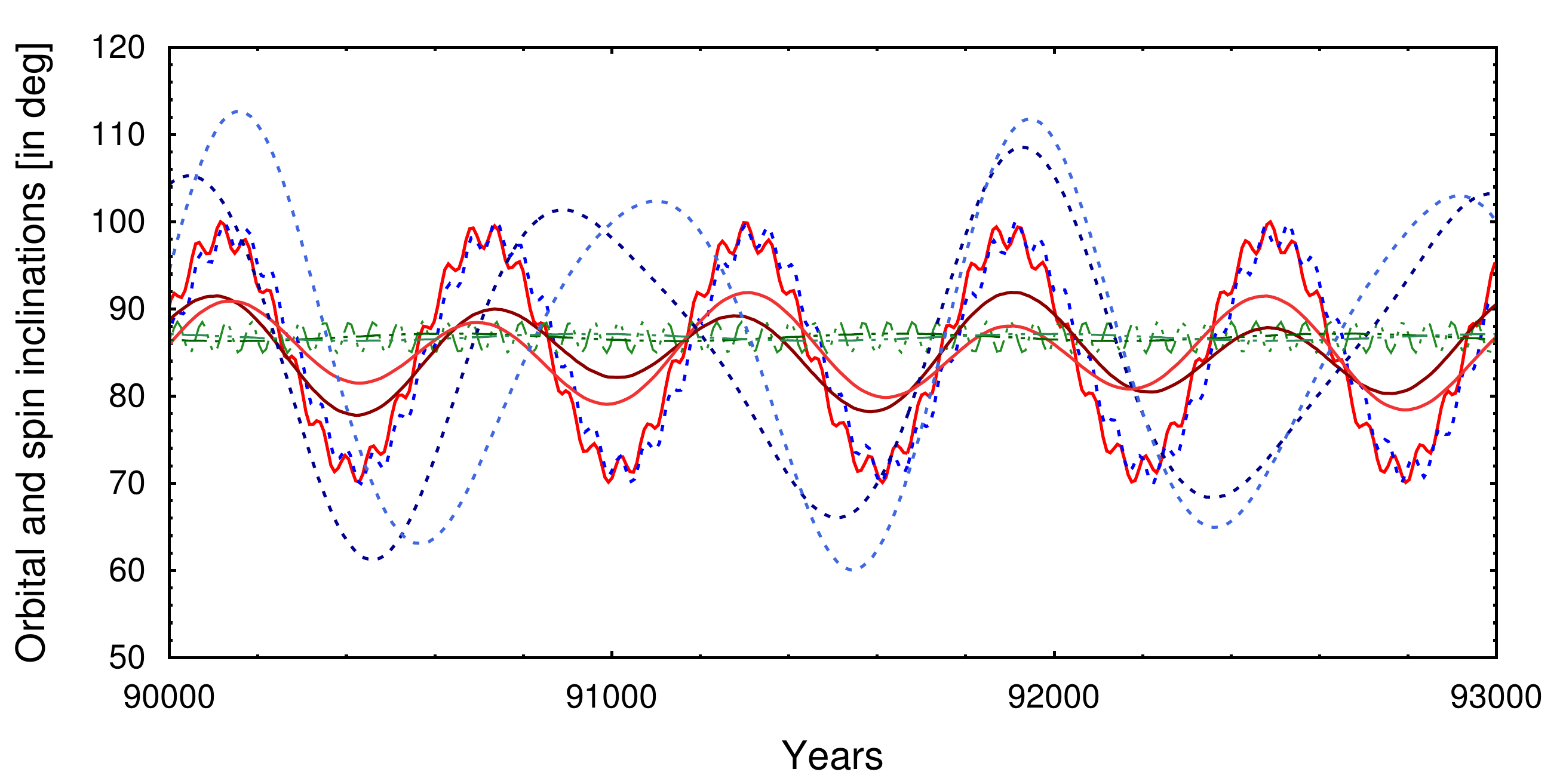}
\includegraphics[width=0.49 \textwidth]{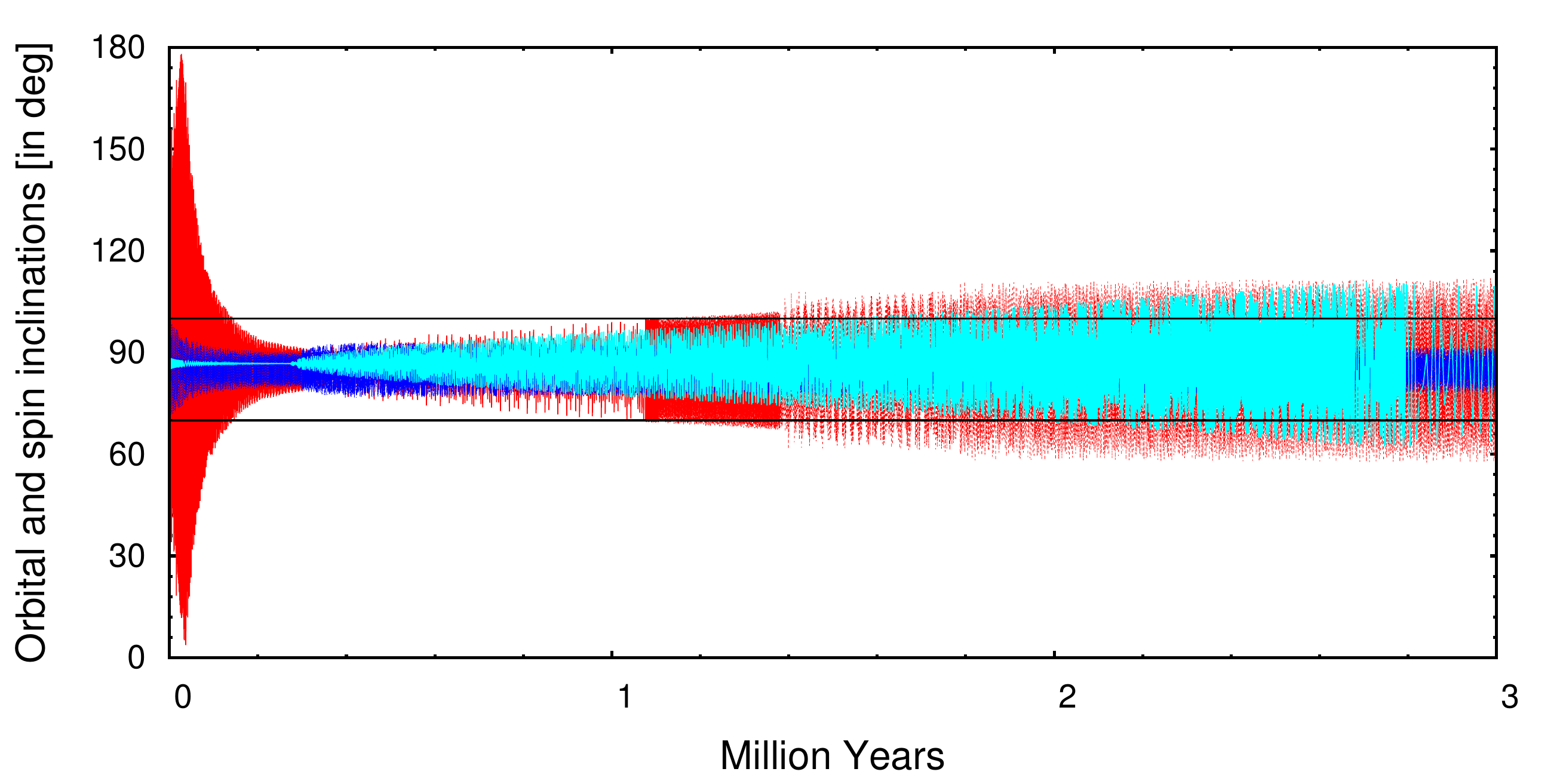}
\caption{ The evolution of the orbital inclination of the innermost
  binary and the orientation of the stellar equators of stars A and B
  during three different integration runs. {\em Left panel:} A
  3000-year-long zoom into the variation of the  angles. Solid lines
  (with three different shades of red) represent the variation of the
  orbital inclination and the equatorial angles of the two stars
  during run `A'. The short-period, small amplitude fluctuations in
  the inclination reflects the precession due to the star C, while
  the longer period, larger amplitude variation is the precession
  caused by the
  star D. Dotted lines (in blue colors) show the same parameters for
  the run `B' (i.e.  inclined spin axes), while the dash-dotted green lines represent run `C', i.e. the originally aligned 3-body model.  {\em Right panel:} 3\,Myr-long evolution of the orientation of the stellar equators. For clarity, we plot only one star for each run. Blue: star A in run `A' (aligned); red: star B in run `B' (inclined); cyan: star A in run `C' (aligned three-body). Horizontal black lines stand for the maximum and minimum values of the orbital inclination of the precessing innermost orbit.}
\label{fig:incltheta} 
\end{center}
\end{figure*}

\begin{figure}
\begin{center}
\includegraphics[width=0.49 \textwidth]{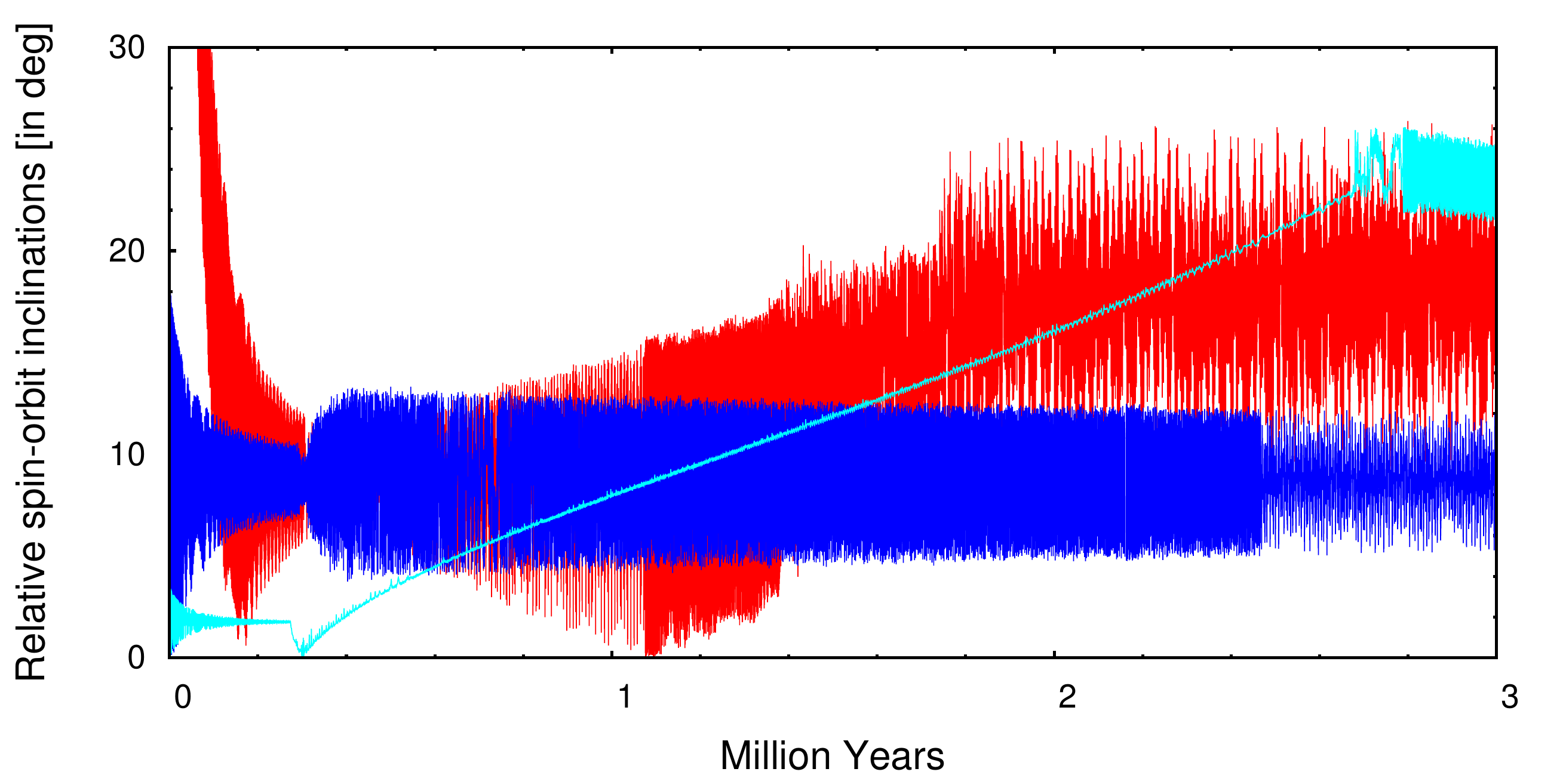}
\caption{ The variation of the mutual inclination angles between the
  innermost orbital plane and stellar equators during 3\,Myr. Blue:
  star A in run `A' (aligned); red: star B in run `B' (inclined);
  cyan: star A in run `C' (aligned three-body). }
\label{fig:inclthetamut}
\end{center}
\end{figure}

\begin{figure}
\begin{center}
\includegraphics[width=0.49 \textwidth]{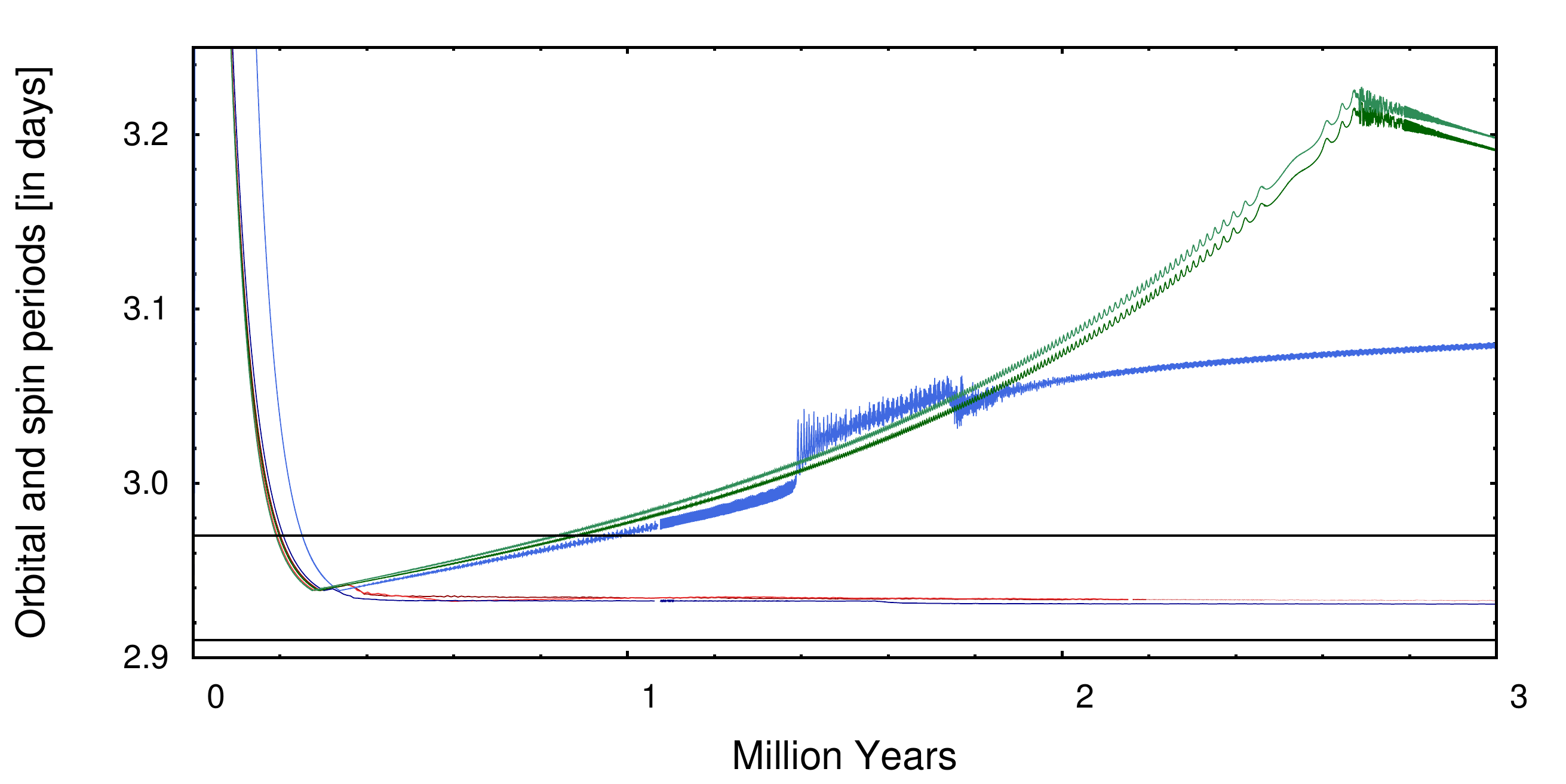}
\caption{ 3\,My-long evolution of the stellar spins of the innermost
  binary members. Two (undistinguishable) red curves stand for the
  spin periods of stars A and B during run `A' (aligned axes). Darker
  and lighter blue lines represent spin
  periods of star A and B, respectively, in run `B' (inclined stellar
  axes). Furthermore, darker and lighter green curves denote the spin
  periods of stars A and B in run `C' (aligned 3-body model). Finally,
  horizontal black lines show the upper and lower values of the
  fluctuating osculating orbital period of the innermost orbit. One
  can see that due to dissipative forces the spin periods quickly
  reach the orbital period. However, in this quasi-synchronized state different spin-orbit coupling and resonances may occur. }
\label{fig:perrot} 
\end{center}
\end{figure}

In Fig.\,\ref{fig:incltheta} we plot the variations of the orbital
inclinations of the innermost orbit and of the equatorial planes of
stars A and B on different time scales. The $\Delta i_1\approx30\degr$
peak-to-peak amplitude, $\approx600$-year-period orbital precession
triggered by the inclined outermost orbit is well visible. The
short-term observational consequences of this additional precession
were briefly discussed a few paragraphs above. As our results
illustrate, this orbital plane precession triggers the  stellar spin
precession with an initially similar amplitude. However, the
equatorial planes of the stars are unable to follow strictly the
orbital plane, as  shown in Fig.\,\ref{fig:inclthetamut}. As a
consequence, the stellar equators no longer remain aligned to the
continuously varying orbital plane. On the other hand,  due to
the dissipative forces the amplitude of the spin precession dumps
quickly in the first few hundred thousand years. Dramatic changes
occur, however, during the later stages of the evolution. The origin of
these changes can be found in some kind of spin-orbit resonances. We
plot the evolution of the stellar spin rates in
Fig.\,\ref{fig:perrot}. As one can see, due to the dissipative forces
the originally sub- or super-synchronous rotation periods quickly relax
to the orbital period. However, in this case various spin-orbit
resonances may occur. As a consequence, the stellar spins can again
desynchronize and, furthermore, large amplitude equatorial plane
precession may also happen. The investigation of these phenomena is
beyond the scope of the present paper; they were studied, e.~g., by
  \citet{correiaetal16}. In the context
of the present paper, we  conclude that the measured low projected
rotational velocities of stars A and B probably offer observational
evidence for the presence of strong spin-orbit coupling. The question
whether the stars have strongly inclined spin axes or  rotate slowly
(or both) cannot be answered at present.\footnote{ We made a period search of the residuals of {\em Kepler} photometry to our photodynamical model for potential signal caused by starspots, and 
have not found any significant periods different from the orbital period and its harmonics.}

\section{Discussion and conclusions}
\label{sec:disc}

The triple system HIP~41431 is remarkable in several respects. First,
it is very compact,  with a 3-tier (3+1) hierarchy  fitting inside the
3.3-au outer orbit. Second, all orbits are close to one plane
(mutual inclinations of 2\fdg2$\pm$0\fdg1  and 21\degr$\pm$16\degr), while the period ratios are
similar (20.17 and 24.4).  The orbits interact dynamically. 

The spatial velocity of this system $(U,V,W) = (8.1, 7.7, -1.4)$ \kms
does not distinguish it from the old disk population and does not
match known kinematical groups of young stars in the solar
neighbourhood. The spectra do not have the lithium 6708\AA ~line or
emissions in H$\alpha$ typical of young stars and no variability
 associated with chromospheric activity or star spots was found in
 the {\em K2} data.  We conclude that this multiple system is not young. 

The  most unusual  characteristic of  this  system is  the slow  axial
rotation of stars comprising  the  short-period inner pair, 
expected to be  tidally synchronzed. However, this apparent paradox might
be caused by  the spin-orbit coupling and resonances triggered by the
dynamically interacting third and fourth stellar companions,  leading
to chaotic rotation.

We  looked for similarly compact hierarchies  in  the
Multiple Star Catalog (MSC) \citep{MSC} . The current version of
the catalog  contains 29 triples with outer
periods $P_\mathrm{out}<150$\,d (not counting the present system).
All MS triples except one have primary components of earlier spectral type than
HIP\,41431 (likely an observational selection
effect). There are only six known triples, however, with the outer
periods shorter than 59\,d. While the absolute dimensions of the
orbits (and, therefore, the orbital periods) are very important
parameters from the point of view of the effectiveness of the tidal
forces and also of the system formation scenarios, the period ratios
are more significant indicators of the strength  of dynamical
interactions between orbits. In this regard, the
period ratios of $\sim$20 found in the two subsystems of HIP\,41431
are far from being extreme. In the small mutual inclination regime, such
period ratios  are well within the stability region of hierarchical
triple stars \citep[see, e.~g.][]{mardlingaarseth01}.

No quadruple systems of 3+1 hierarchy as compact as
HIP~41431 were known previously. However,  there are at least three
 compact triple systems with short outer periods  among the {\em
   Kepler}'s prime misson EBs where the systematic residuals of the
 four-year-long ETV data might indicate the presence of a fourth
 component \citep{Borkovits2016}. 

Our work has contributed an interesting system that challenges the
theories of star formation.
Compact and coplanar hierarchical stellar systems like HIP~41431 
are probably a  result of  migration in massive disks that are present at
the time of star formation. The first two stellar embryos condense from gas,
accrete mass, and migrate inward, while outer components condense
later in the same accretion flow (e.~g. by disc fragmentation) and, in their turn,
migrate inward. No other scenario can plausibly explain the origin
of such well-organized, planetary-like hierarchies. However, further discussion
of formation mechanisms is beyond the scope of this paper.

\section*{Acknowledgments}

We thank the Referee, D.~Gies, for useful suggestions and corrections. 
T.\,B. acknowledges the financial support of the Hungarian National Research, 
Development and Innovation Office -- NKFIH Grants OTKA K-113117 and KH-130372.
L. M. was supported by the Premium Postdoctoral Program of the Hungarian Academy of Sciences. The research leading to these results has received funding from the LP2017-8 Lend\"ulet grant of the Hungarian Academy of Sciences. 

We  used the  Simbad  service  operated by  the  Centre des  Donn\'ees
Stellaires (Strasbourg,  France) and the ESO  Science Archive Facility
services (data  obtained under request number 396301).   This work has
made use  of data  from the European  Space Agency (ESA)  mission {\it
  Gaia}\footnote{\url{https://www.cosmos.esa.int/gaia}},  processed  by  the
{\it   Gaia}   Data   Processing   and  Analysis   Consortium   (DPAC,
\url{https://www.cosmos.esa.int/web/gaia/dpac/consortium}).  Funding for the
DPAC  has been provided  by national  institutions, in  particular the
institutions participating in the {\it Gaia} Multilateral Agreement.

This paper includes data collected by the {\em K2} mission. Funding for the {\em K2} mission is provided by the NASA Science Mission directorate. Some of the data presented in this paper were obtained from the Mikulski Archive for Space Telescopes (MAST). STScI is operated by the Association of Universities for Research in Astronomy, Inc., under NASA contract NAS5-26555. Support for MAST for non-HST data is provided by the NASA Office of Space Science via grant NNX09AF08G and by other grants and contracts.





\appendix
\onecolumn

\section{Some details of the numerical integrator for modelling 2+1+1 hierarchies}
\label{app:numint}

As it was mentioned above, the numerical integrator which was used in
the spectro-photodynamical code is an upgraded version of the 3-body
integrator  described in \citet{borkovitsetal04}. Further details
of the practical implementation of a numerical integrator coupled to
the  lightcurve emulator were discussed in the appendix of \citet{Borkovits2019}. Here we discuss the additional modifications introduced into the code to handle quadruple systems with 3+1 hierarchy.

Similar to the previous hierarchical triple star case, the Jacobian
vector formalism is consistently used. In order to describe the motion
of the fourth body and its effect on the inner three stars, now we introduce the third Jacobian vector which points to the outermost component (star D) from the centre of mass of the inner triple subsystem (stars A, B, and C).

Let us denote by $\vec{r}_i$ the barycentric radius vector of the
component $i$ and by $\vec{r}_{ij}=\vec{r}_j-\vec{r}_i$ the vector between components $j$ and $i$. Then, the first three Jacobian vectors are as follows:
\begin{eqnarray}
\vec{\rho}_1&=&\vec{r}_{12}, \label{ro1def5} \\
\vec{\rho}_2&=&\vec{r}_{3}-\frac{m_1}{m_{12}}\vec{r}_1-\frac{m_2}{m_{12}}\vec{r}_2=\vec{r}_{13}-\frac{m_2}{m_{12}}\vec{r}_{12}=\vec{r}_{23}+\frac{m_1}{m_{12}}\vec{r}_{12}, \label{r2def5} \\
\vec{\rho}_3&=&\vec{r}_{4}-\frac{m_1}{m_{123}}\vec{r}_1-\frac{m_2}{m_{123}}\vec{r}_2-\frac{m_3}{m_{123}}\vec{r}_3,
\label{ro3def5}
\end{eqnarray}
while the mutual distances between the components are:
\begin{eqnarray}
\vec{r}_{12}&=&\vec\rho_1,  \\
\vec{r}_{13}&=&\vec\rho_2+\frac{m_2}{m_{12}}\vec\rho_1 \\
\vec{r}_{14}&=&\vec\rho_3+\frac{m_2}{m_{12}}\vec\rho_1+\frac{m_3}{m_{123}}\vec\rho_2 \\
\vec{r}_{23}&=&\vec\rho_2-\frac{m_1}{m_{12}}\vec\rho_1 \\
\vec{r}_{24}&=&\vec\rho_3-\frac{m_1}{m_{12}}\vec\rho_1+\frac{m_3}{m_{123}}\vec\rho_2 \\
\vec{r}_{34}&=&\vec\rho_3-\frac{m_{12}}{m_{123}}\vec\rho_2.
\end{eqnarray}
Then, the point-mass ($U$), tidal ($T$), and rotational ($R$) components of total potential take the following forms:
\begin{equation}
U=\frac{Gm_1m_2}{r_{12}}+\frac{Gm_1m_3}{r_{13}}+\frac{Gm_2m_3}{r_{23}}+\frac{Gm_1m_4}{r_{14}}+\frac{Gm_2m_4}{r_{24}}+\frac{Gm_3m_4}{r_{34}},
\label{tomegpontpoti}
\end{equation}
\begin{equation}
T_\mathrm{12}=\frac{Gm_1m_2}{r_{12}}\sum_{j=2}^4\left\{\frac{m_2}{m_1}2k_j^{(1)}\left(\frac{R_1}{r_{12}}\right)^j\left(\frac{R_1}{r_{\mathrm{d}_1}}\right)^{j+1}P_j(\lambda_1)+\frac{m_1}{m_2}2k_j^{(2)}\left(\frac{R_2}{r_{12}}\right)^j\left(\frac{R_2}{r_{\mathrm{d}_2}}\right)^{j+1}P_j(\lambda_2)\right\},
\label{arapalypoti125}
\end{equation}
\begin{equation}
T_\mathrm{12out}=\sum_{i=1}^2\sum_{\ell=3}^4\frac{Gm_im_\ell}{r_{i\ell}}\frac{m_{3-i}}{m_i}2k_2^{(i)}\left(\frac{R_i}{r_{i\ell}}\right)^2\left(\frac{R_i}{r_{12}}\right)^3P_2(\lambda_{i\ell})+\frac{Gm_1m_2}{r_{12}}\sum_{i=1}^2\sum_{\ell=3}^4\frac{m_\ell}{m_i}2k_2^{(i)}\left(\frac{R_i}{r_{12}}\right)^2\left(\frac{R_i}{r_{i\ell}}\right)^3P_2(\lambda_{i\ell}),
\label{arapalypoti12x5}
\end{equation}
and, furthermore,
\begin{equation}
R_\mathrm{12}=\frac{Gm_im_\ell}{r_{i\ell}}\sum_{i=1}^2\sum_{\stackrel{\ell=1}{\ell\neq i}}^4\left\{\frac{k_2^{(i)}R_i^5}{Gm_i}\left[\frac{\omega_{z'_i}^2}{3r_{i\ell}^2}-\frac{(\vec{r}_{i\ell}\cdot\vec{\omega}_{z'_i})^2}{r_{i\ell}^4}\right]\right\}.
\label{forgaspoti125}
\end{equation}
(These expressions were deduced with the 2+1+1 case generalization of the Eqs.\,(10--13)
of \citealp{borkovitsetal04}, based on the treatment of \citealt{kopal78}.)
The tidal and rotational terms are calculated only for stars A and B
(denoted here by indices 1 and 2), i.~e. for the members of the
innermost binary. In these expressions, $R_i$ denotes the radius of
the $i$-th star, $k_\mathrm{j}^{(i)}$ stands for the $j$-th apsidal
motion constant of the $i$-th star (practically only $k_2$-s were
used). Furthermore, in the $T_\mathrm{12}$ term which describes the
mutual interaction between the close binary members,
$r_{\mathrm{d}_i}$ is the distance between the two stars, taken into
account the tidal lag time of the component $i$, and $\lambda_i$ denotes the direction cosine between the radius vector and the tidal bulge of the $i$-th star. For the non-dissipative case, which was used for the spectro-photodynamical runs, $d_1=d_2=r_{12}$ and $\lambda_1=\lambda_2=1$.
The terms $T_{12out}$ give the tidal contributions of stars C and D to the motion of the innermost binary. Finally, in the last term ($R_\mathrm{12}$), which describes the contributions of the rotational oblateness of stars A and B, $\vec{\omega}_{z'_i}$ stands for the uni-axial spin angular momentum vector of the $i$-th component.

With the use of these potential terms, the equations of the motions to be integrated take the following form:
\begin{eqnarray}
\ddot{\vec{\rho}}_1&=&-\frac{Gm_{12}}{\rho_1^3}\vec{\rho}_1+Gm_3\left(\frac{\vec{r}_{23}}{r_{23}^3}-\frac{\vec{r}_{13}}{r_{13}^3}\right)+Gm_4\left(\frac{\vec{r}_{24}}{r_{24}^3}-\frac{\vec{r}_{14}}{r_{14}^3}\right) \nonumber \\
&&-\frac{Gm_{12}}{\rho_1^3}\left\{\sum_{i=1}^2\left\{\sum_{j=2}^4\frac{m_{3-i}}{m_i}2(j+1)k^{(i)}_j\left(\frac{R_i}{\rho_1}\right)^j\left(\frac{R_i}{r_{\mathrm{d}_i}}\right)^{j+1}\vec{\cal{P}}_j(\lambda_i)+\frac{k^{(i)}_2R_i^5}{Gm_i}\left\{\left[\frac{\omega_{z'_i}^2}{\rho_1^2}-5\frac{(\vec{\rho}_1\cdot\vec{\omega}_{z'_i})^2}{\rho_1^4}\right]\vec{\rho}_1+\frac{2\vec{\rho}_1\cdot\vec{\omega}_{z'_i}}{\rho_1^2}\vec{\omega}_{z'_i}\right\}\right\}\right\} \nonumber \\
&&+\sum_{\ell=3}^4m_\ell\sum_{i=1}^2(-1)^i\frac{k^{(i)}_2R_i^5}{m_ir_{i\ell}^5}\left\{\left[\omega_{z'_i}^2-5\frac{(\vec{r}_{i\ell}\cdot\vec{\omega}_{z'_i})^2}{r_{i\ell}^2}\right]\vec{r}_{i\ell}+2(\vec{r}_{i\ell}\cdot\vec{\omega}_{z'_i})\vec{\omega}_{z'_i}\right\} \nonumber \\
&&+\sum_{\ell=3}^43Gm_\ell\sum_{i=1}^2\left\{(-1)^i\frac{m_{3-i}k^{(i)}_2R_i^5}{m_ir_{i\ell}^5\rho_1^3}\left\{\left[5\frac{(\vec{r}_{i\ell}\cdot\vec{\rho}_1)^2}{\rho_1^2r_{i\ell}^2}-1\right]\vec{r}_{i\ell}-2\frac{\vec{r}_{i\ell}\cdot\vec{\rho}_1}{\rho_1^2}\vec{\rho}_1\right\}\right. \nonumber \\
&&\left.-\frac{m_{12}k^{(i)}_2R_i^5}{m_ir_{i\ell}^3\rho_1^5}\left\{\left[5\frac{(\vec{r}_{i\ell}\cdot\vec{\rho}_1)^2}{\rho_1^2r_{i\ell}^2}-1\right]\vec{\rho}_1-2\frac{\vec{r}_{i\ell}\cdot\vec{\rho}_1}{r_{i\ell}^2}\vec{r}_{i\ell}\right\}\right\},
\label{mozgro1}
\end{eqnarray}
\begin{eqnarray}
\ddot{\vec\rho}_2&=&-\frac{Gm_{123}}{m_{12}}\left(\frac{m_1}{r_{13}^3}\vec{r}_{13}+\frac{m_2}{r_{23}^3}\vec{r}_{23}\right)+\frac{Gm_4}{m_{12}}\left(\frac{m_{12}}{r_{34}^3}\vec{r}_{34}-\frac{m_1}{r_{14}^3}\vec{r}_{14}--\frac{m_2}{r_{24}^3}\vec{r}_{24}\right) \nonumber \\
&&-\frac{m_{123}}{m_{12}}\left\{\sum_{i=1}^2\frac{k_2^{(i)}R_i^5}{r_{i3}^5}\left\{\left\{\left[\omega_{z'_i}^2-5\frac{(\vec{r}_{i3}\cdot\omega_{z'_i})^2}{r_{i3}^2}\right]\vec{r}_{i3}+2(\vec{r}_{i3}\cdot\vec{\omega_{z'_i}})\vec{\omega_{z'_i}}\right\}+\frac{3Gm_{3-i}}{\rho_1^3}\left\{\left[5\frac{(\vec\rho_1\cdot\vec{r}_{i3})^2}{\rho_1^2r_{i3}^2}-1\right]\vec{r}_{i3}-2\frac{\vec\rho_1\cdot\vec{r}_{i3}}{\rho_1^2}\vec\rho_1\right\}\right\}\right\} \nonumber \\
&&-\frac{m_4}{m_{12}}\left\{\sum_{i=1}^2\frac{k_2^{(i)}R_i^5}{r_{i4}^5}\left\{\left\{\left[\omega_{z'_i}^2-5\frac{(\vec{r}_{i4}\cdot\omega_{z'_i})^2}{r_{i4}^2}\right]\vec{r}_{i4}+2(\vec{r}_{i4}\cdot\vec{\omega_{z'_i}})\vec{\omega_{z'_i}}\right\}+\frac{3Gm_{3-i}}{\rho_1^3}\left\{\left[5\frac{(\vec\rho_1\cdot\vec{r}_{i4})^2}{\rho_1^2r_{i4}^2}-1\right]\vec{r}_{i4}-2\frac{\vec\rho_1\cdot\vec{r}_{i4}}{\rho_1^2}\vec\rho_1\right\}\right\}\right\}. \nonumber \\
\label{mozgro2}
\end{eqnarray}
\begin{eqnarray}
\ddot{\vec\rho}_3&=&-\frac{m_{1234}}{m_{123}}\left\{\frac{Gm_1}{r_{14}^3}\vec{r}_{14}+\frac{Gm_2}{r_{24}^3}\vec{r}_{24}+\frac{Gm_3}{r_{34}^3}\vec{r}_{34}\right. \nonumber \\
&&\left.+\sum_{i=1}^2\frac{k_2^{(i)}R_i^5}{r_{i4}^5}\left\{\left\{\left[\omega_{z'_i}^2-5\frac{(\vec{r}_{i4}\cdot\omega_{z'_i})^2}{r_{i4}^2}\right]\vec{r}_{i4}+2(\vec{r}_{i4}\cdot\vec{\omega_{z'_i}})\vec{\omega_{z'_1}}\right\}+\frac{3Gm_{3-i}}{\rho_1^3}\left\{\left[5\frac{(\vec\rho_1\cdot\vec{r}_{i4})^2}{\rho_1^2r_{i4}^2}-1\right]\vec{r}_{i4}-2\frac{\vec\rho_1\cdot\vec{r}_{i4}}{\rho_1^2}\vec\rho_1\right\}\right\}\right\}, \nonumber \\
\label{mozgro3}
\end{eqnarray}
where
\begin{eqnarray}
\vec{\cal{P}}_2(\lambda_i)&=&P_2(\lambda_i)\vec{\rho}_1+\frac{\lambda_i}{\rho_1r_{\mathrm{d}_i}}(\vec{r}_{\mathrm{d}_i}\times\vec{\rho}_1)\times\vec{\rho}_1=\frac{1}{2}\left[5\frac{(\vec{\rho}_1\cdot\vec{r}_{\mathrm{d}_i})^2}{\rho_1^2r_{\mathrm{d}_i}^2}-1\right]\vec{\rho}_1-\frac{\vec{\rho}_1\cdot\vec{r}_{\mathrm{d}_i}}{r_{\mathrm{d}_i}^2}\vec{r}_{\mathrm{d}_i}, \label{Eq:P2}\\
\vec{\cal{P}}_3(\lambda_i)&=&P_3(\lambda_i)\vec{\rho}_1+\frac{3}{2}\frac{5\lambda^2_i-1}{\rho_1r_{\mathrm{d}_i}}(\vec{r}_{\mathrm{d}_i}\times\vec{\rho}_1)\times\vec{\rho}_1=\left[10\frac{(\vec{\rho}_1\cdot\vec{r}_{\mathrm{d}_i})^3}{\rho_1^3r_{\mathrm{d}_i}^3}-3\frac{\vec{\rho}_1\cdot\vec{r}_{\mathrm{d}_i}}{\rho_1r_{\mathrm{d}_i}}\right]\vec{\rho}_1-\frac{3}{2}\left[5\frac{(\vec{\rho}_1\cdot\vec{r}_{\mathrm{d}_i})^2}{\rho_1r_{\mathrm{d}_i}^3}-\frac{\rho_1}{r_{\mathrm{d}_i}}\right]\vec{r}_{\mathrm{d}_i}, \\
\vec{\cal{P}}_4(\lambda_i)&=&P_4(\lambda_i)\vec{\rho}_1+\frac{5}{8}\frac{7\lambda^3_i-3\lambda_i}{\rho_1r_{\mathrm{d}_i}}(\vec{r}_{\mathrm{d}_i}\times\vec{\rho}_1)\times\vec{\rho}_1=\frac{1}{8}\!\!\left[\!70\frac{(\vec{\rho}_1\cdot\vec{r}_{\mathrm{d}_i})^4}{\rho_1^4r_{\mathrm{d}_i}^4}\!\!-\!45\frac{(\vec{\rho}_1\cdot\vec{r}_{\mathrm{d}_i})^2}{\rho_1^2r^2_{\mathrm{d}_i}}\!\!+\!3\!\right]\!\vec{\rho}_1\!\!-\!\frac{5}{8}\!\left[\!7\!\frac{(\vec{\rho}_1\cdot\vec{r}_{\mathrm{d}_i})^3}{\rho_1^2r_{\mathrm{d}_i}^4}\!\!-\!3\!\frac{\vec{\rho}_1\cdot\vec{r}_{\mathrm{d}_i}}{r_{\mathrm{d}_i}^2}\!\right]\!\!\vec{r}_{\mathrm{d}_i}. \nonumber \\
\label{Eq:P4}
\end{eqnarray}
Note, however, that for non-dissipative cases $\vec{\rho}_1=\pm\vec{r}_{\mathrm{d}_{1,2}}$ and thus, Eqs.~(\ref{Eq:P2}--\ref{Eq:P4}) reduce simply to 
\begin{equation}
\vec{\cal{P}}_j(\lambda_i)=\vec{\rho}_1.
\end{equation}

The spin angular momentum vectors of stars A and B ($\vec{\omega_{z'_{1,2}}}$) may take any arbitrary orientations and magnitude. Their evolution is also numerically integrated simultaneously via the Eulerian equations of rotation. The corresponding expressions are given in Eqs.\,(B.11--B.13) of \citet{borkovitsetal04} and, therefore, we do not repeat them here.

\section{Supplementary material}

\subsection{Times of eclipsing minima for ETV studies}

In this subsection we tabulate the times of eclipsing minima of HIP\,41431. The full list is available online only and, for reader's convenience, it is provided in machine readable format.

\begin{table*}
\caption{Times of minima of HIP\, 41431 (EPIC\,212096658). }
 \label{Tab:EPIC_212096658_ToM}
\begin{tabular}{@{}lrllrllrl}
\hline
BJD & Cycle  & std. dev. & BJD & Cycle  & std. dev. & BJD & Cycle  & std. dev. \\ 
$-2\,400\,000$ & no. &   \multicolumn{1}{c}{$(d)$} & $-2\,400\,000$ & no. &   \multicolumn{1}{c}{$(d)$} & $-2\,400\,000$ & no. &   \multicolumn{1}{c}{$(d)$} \\ 
\hline
57140.546616 &   -1.0 & 0.000745 & 57212.416959 &   23.5 & 0.000101 & 58165.600764 &  348.5 & 0.000344 \\
57142.030814 &   -0.5 & 0.000244 & 57213.863215 &   24.0 & 0.000159 & 58167.071306 &  349.0 & 0.000334 \\
57143.477973 &    0.0 & 0.000170 & 58096.686632 &  325.0 & 0.000451 & 58168.532472 &  349.5 & 0.000294 \\
57144.962745 &    0.5 & 0.000078 & 58098.149278 &  325.5 & 0.000653 & 58170.002961 &  350.0 & 0.000886 \\
57146.409726 &    1.0 & 0.000214 & 58099.617845 &  326.0 & 0.000084 & 58171.464520 &  350.5 & 0.000783 \\
57147.894165 &    1.5 & 0.000074 & 58101.080692 &  326.5 & 0.001293 & 58172.935285 &  351.0 & 0.000819 \\
57149.340890 &    2.0 & 0.000046 & 58102.549156 &  327.0 & 0.001777 & 58174.396859 &  351.5 & 0.003913 \\
57150.825786 &    2.5 & 0.000077 & 58104.012053 &  327.5 & 0.000132 & 58252.136475 &  378.0 & 0.001828 \\
57152.272733 &    3.0 & 0.000028 & 58105.480644 &  328.0 & 0.000211 & 58253.596266 &  378.5 & 0.000592 \\
57153.757104 &    3.5 & 0.000192 & 58106.943575 &  328.5 & 0.000624 & 58255.071344 &  379.0 & 0.000532 \\
57155.204030 &    4.0 & 0.000019 & 58108.412067 &  329.0 & 0.000687 & 58256.529671 &  379.5 & 0.000108 \\
57156.688796 &    4.5 & 0.000063 & 58109.875621 &  329.5 & 0.000159 & 58258.004748 &  380.0 & 0.003455 \\
57158.135547 &    5.0 & 0.000160 & 58111.343495 &  330.0 & 0.001087 & 58259.462186 &  380.5 & 0.000881 \\
57159.620606 &    5.5 & 0.000113 & 58112.807034 &  330.5 & 0.000873 & 58260.937413 &  381.0 & 0.000093 \\
57161.066880 &    6.0 & 0.000127 & 58114.275684 &  331.0 & 0.000479 & 58262.394033 &  381.5 & 0.000125 \\
57162.552206 &    6.5 & 0.000171 & 58115.739490 &  331.5 & 0.000912 & 58263.869333 &  382.0 & 0.001153 \\
57163.998782 &    7.0 & 0.000354 & 58117.208752 &  332.0 & 0.000199 & 58265.325674 &  382.5 & 0.001305 \\
57165.483852 &    7.5 & 0.000130 & 58118.672846 &  332.5 & 0.001196 & 58266.801169 &  383.0 & 0.000853 \\
57166.930611 &    8.0 & 0.000351 & 58120.143383 &  333.0 & 0.000865 & 58268.257329 &  383.5 & 0.000043 \\
57168.415775 &    8.5 & 0.000106 & 58121.607314 &  333.5 & 0.000263 & 58269.732583 &  384.0 & 0.002651 \\
57169.863229 &    9.0 & 0.000172 & 58123.079612 &  334.0 & 0.000102 & 58271.188833 &  384.5 & 0.000376 \\
57171.348066 &    9.5 & 0.000161 & 58124.543101 &  334.5 & 0.000998 & 58272.664065 &  385.0 & 0.000550 \\
57172.796124 &   10.0 & 0.000167 & 58126.015778 &  335.0 & 0.000776 & 58274.120305 &  385.5 & 0.000204 \\
57174.281235 &   10.5 & 0.000057 & 58127.478111 &  335.5 & 0.001329 & 58275.595426 &  386.0 & 0.002417 \\
57175.730707 &   11.0 & 0.000100 & 58128.949092 &  336.0 & 0.000484 & 58277.051804 &  386.5 & 0.002369 \\
57177.216441 &   11.5 & 0.000347 & 58130.412798 &  336.5 & 0.000113 & 58278.526897 &  387.0 & 0.000123 \\
57178.666657 &   12.0 & 0.000198 & 58131.883669 &  337.0 & 0.002043 & 58279.983353 &  387.5 & 0.000183 \\
57180.153887 &   12.5 & 0.000225 & 58133.348268 &  337.5 & 0.000071 & 58281.458184 &  388.0 & 0.000686 \\
57181.603000 &   13.0 & 0.000106 & 58134.819223 &  338.0 & 0.000474 & 58282.914900 &  388.5 & 0.007737 \\
57183.090425 &   13.5 & 0.000073 & 58136.282722 &  338.5 & 0.000920 & 58284.389716 &  389.0 & 0.000481 \\
57184.537118 &   14.0 & 0.000056 & 58137.753601 &  339.0 & 0.009431 & 58285.846670 &  389.5 & 0.002594 \\
57186.024259 &   14.5 & 0.000231 & 58139.215825 &  339.5 & 0.000868 & 58287.321398 &  390.0 & 0.000061 \\
57187.471259 &   15.0 & 0.000578 & 58140.686940 &  340.0 & 0.000300 & 58288.778231 &  390.5 & 0.000360 \\
57188.958197 &   15.5 & 0.000172 & 58142.148125 &  340.5 & 0.000455 & 58290.253313 &  391.0 & 0.000767 \\
57190.406446 &   16.0 & 0.000021 & 58143.619238 &  341.0 & 0.011193 & 58291.710867 &  391.5 & 0.004955 \\
57191.892233 &   16.5 & 0.000169 & 58145.079962 &  341.5 & 0.000609 & 58293.186097 &  392.0 & 0.000045 \\
57193.340629 &   17.0 & 0.000077 & 58146.551272 &  342.0 & 0.001466 & 58294.643694 &  392.5 & 0.036326 \\
57194.825643 &   17.5 & 0.000281 & 58148.011626 &  342.5 & 0.000309 & 58296.120404 &  393.0 & 0.001080 \\
57196.273525 &   18.0 & 0.000091 & 58149.482848 &  343.0 & 0.008323 & 58297.577843 &  393.5 & 0.000057 \\
57197.758109 &   18.5 & 0.000103 & 58150.943127 &  343.5 & 0.000941 & 58299.055921 &  394.0 & 0.000481 \\
57199.205659 &   19.0 & 0.000257 & 58152.414341 &  344.0 & 0.000714 & 58300.513091 &  394.5 & 0.000900 \\
57200.690248 &   19.5 & 0.000385 & 58153.874627 &  344.5 & 0.000127 & 58301.992113 &  395.0 & 0.003337 \\
57202.137571 &   20.0 & 0.000208 & 58155.345636 &  345.0 & 0.002435 & 58498.506621 &  462.0 & 0.000031 \\
57203.622066 &   20.5 & 0.000147 & 58156.806057 &  345.5 & 0.001051 & 58542.496480 &  477.0 & 0.000029 \\
57205.069162 &   21.0 & 0.000148 & 58158.276984 &  346.0 & 0.001531 & 58548.366974 &  479.0 & 0.000031 \\
57206.553610 &   21.5 & 0.000239 & 58159.737593 &  346.5 & 0.000636 & 58567.409913 &  485.5 & 0.000022 \\
57208.000524 &   22.0 & 0.000127 & 58161.208348 &  347.0 & 0.000067 & 58570.341545 &  486.5 & 0.000022 \\
57209.485395 &   22.5 & 0.000282 & 58162.669118 &  347.5 & 0.001053 & 58592.350572 &  494.0 & 0.000022 \\
57210.931931 &   23.0 & 0.000050 & 58164.139718 &  348.0 & 0.000398 &  &   &  \\
\hline
\end{tabular}

{\em Notes.} Integer and half-integer cycle numbers refer to primary and secondary eclipses, respectively. Most of the eclipses (cycle nos. $-1.0$ to 395.0) were observed by {\em Kepler} spacecraft. The last six eclipses were observed at Baja Astronomical Observatory.
\end{table*}

\subsection{Radial velocity data} 

The columns give the BJD date of observation, the RVs of the components A, B, and C in \kms, their residuals to the spectro-photodynamical model, and the instrument code (see Sect.~\ref{subsec:spectroscopy}.). In fitting the RVs, we adopt the errors of 2.0 \kms for CfA, 0.5 \kms for TRES, UVES, VUES, and CHIRON. For reader's convenience, the full, online available list is provided in machine readable format.

\begin{table*}
\caption{Radial velocity data of the three components of HIP\, 41431. }
 \label{Tab:EPIC_212096658ABC_RV}
\begin{tabular}{@{}lrrrrrrl}
\hline
BJD & $\mathrm{RV}_A$  & $\Delta\mathrm{RV}_A$ & $\mathrm{RV}_B$  & $\Delta\mathrm{RV}_B$ & $\mathrm{RV}_C$  & $\Delta\mathrm{RV}_C$ & instr. \\ 
$-2\,400\,000$ & \multicolumn{2}{r}{(km\,s$^{-1})$} & \multicolumn{2}{c}{(km\,s$^{-1})$} & \multicolumn{2}{c}{(km\,s$^{-1})$} &  \\ 
\hline
51295.7154  & -92.50   & +3.08&    42.20  & -0.38 &   28.90 & +0.23 &  CfA  \\     
51325.6699  & -43.20   & -1.63&    66.40  & +1.49 &  -43.20 & +1.09 &  CfA  \\     
51471.9757  & -103.90  & +0.60&    55.50  & +0.77 &   31.00 & +0.23 &  CfA  \\     
51503.0469  &  77.40   & +1.06&   -47.80  & +0.99 &  -47.80 & -2.24 &  CfA  \\     
51505.0016  &  29.10   & -1.44&   8.20    & +2.61 &  -52.90 & +0.19 &  CfA  \\  
51506.9915  & -54.50   & +0.78&    97.70  & -1.27 &  -60.50 & -1.44 &  CfA  \\     
51539.9870  & -46.30   & -1.74&   6.20    & +3.25 &   22.40 & -0.57 &  CfA  \\     
51566.7523  & 56.70    & -2.64&   -13.30  & +2.19 &  -60.00 & +0.38 &  CfA  \\     
51568.7912  & -59.00   & -2.95&   102.10  & -0.06 &  -58.90 & +1.94 &  CfA  \\     
51595.8217  & -30.20   & -0.91&   -12.30  & +4.73 &   27.90 & +0.44 &  CfA  \\     
51620.8426  & 6.50     & +3.17&    21.30  & -4.08 &  -45.30 & -0.20 &  CfA  \\  	
51622.7819  & 89.20    & -0.61&   -52.60  & +2.12 &  -52.50 & +0.01 &  CfA  \\  	
51623.7550  &	17.60  & +5.07&    21.10  & -6.01 &  -56.20 & -0.46 &  CfA  \\  	
51647.6866  &  -92.60  & +2.98&    46.60  & +0.70 &   32.30 & +0.99 &  CfA  \\  	
51648.6177  &  -34.00  & -2.48&   -14.90  & +4.44 &   31.30 & -0.17 &  CfA  \\  	
51649.6840  &	40.60  & +1.42&   -91.80  & -0.80 &   32.30 & +0.96 &  CfA  \\     
51652.6459  &	37.40  & +0.45&   -88.50  & -1.59 &   28.80 & -0.74 &  CfA  \\      
51684.6315  &	98.40  & -2.04&   -60.00  & -1.84 &  -60.10 & +0.30 &  CfA  \\  	
51686.6514  &	-1.40  & -3.51&    44.90  & +2.58 &  -61.10 & -0.05 &  CfA  \\  	
51834.9496  &  -28.10  & +3.43&   -16.20  & -3.19 &   18.30 & -1.09 &  CfA  \\  	
51856.0416  &  -66.20  & +1.16&    89.20  & -0.46 &  -45.50 & -0.21 &  CfA  \\  	
51883.8218  &	40.50  & -1.58&    -98.90 & +2.12 &   29.10 & -0.37 &  CfA  \\  	
51884.0374  &	50.10  & +0.29&   -107.80 & +1.07 &   29.70 & +0.23 &  CfA  \\  	
51884.8582  &  -42.80  & +2.61&    -15.20 & -3.11 &   28.20 & -1.12 &  CfA  \\  	
51885.0429  &  -71.10  & +3.47&     15.10 & -2.48 &   30.10 & +0.84 &  CfA  \\  	
51885.8325  &  -87.00  & +0.11&     32.10 & +1.45 &   29.80 & +0.90 &  CfA  \\      
51886.0330  &  -60.70  & -1.33&     6.00  & +3.41 &   29.30 & +0.52 &  CfA  \\      
51886.8373  &	47.30  & -0.21&   -103.20 & +2.19 &   27.30 & -0.90 &  CfA  \\      
51886.9558  &	51.00  & +0.60&   -107.00 & +1.23 &   28.20 & +0.09 &  CfA  \\      
51887.8637  &  -53.70  & +3.00&     -2.30 & -3.63 &   26.70 & -0.57 &  CfA  \\      
51887.9953  &  -73.70  & +2.77&     18.90 & -2.64 &   26.40 & -0.74 &  CfA  \\      
51888.8238  &  -80.90  & -2.46&     26.40 & +1.94 &   26.20 & -0.00 &  CfA  \\  	
51889.9595  &	52.50  & +1.16&   -104.60 & +1.25 &   24.70 & -0.04 &  CfA  \\  	
51890.8406  &  -57.60  & +4.43&     9.20  & -1.34 &   23.70 & +0.24 &  CfA  \\  	
51891.0314  &  -83.90  & +3.64&     34.90 & -1.84 &   22.20 & -0.96 &  CfA  \\  	
51937.7043  &  -47.10  & +5.40&    -5.50  & -3.63 &   25.50 & +2.18 &  CfA  \\  	
51938.7805  &  -69.30  & -2.90&    13.20  & +3.32 &   26.20 & +0.68 &  CfA  \\     
51942.7051  &	48.50  & +0.12&  -109.30  & +0.77 &   30.30 & +1.73 &  CfA  \\      
51944.6573  &  -67.40  & -1.37&     8.90  & +2.32 &   29.50 & +1.51 &  CfA  \\      
51962.6696  &	19.90  & -0.81&   -42.10  & +2.99 &   -9.00 & -1.00 &  CfA  \\      
51967.6505  &  -82.00  & +0.70&    76.50  & -0.25 &  -24.30 & +0.14 &  CfA  \\      
51997.6097  &  -44.40  & -4.04&   -17.00  & +2.01 &   23.40 & -1.03 &  CfA  \\      
52007.6703  &	20.20  & +5.58&   -76.40  & -2.77 &   20.30 & -2.41 &  CfA  \\      
52008.7070  & -107.60  & +0.34&    52.00  & -0.40 &   21.50 & +0.35 &  CfA  \\     
52009.6658  &	11.50  & -2.85&   -67.00  & +3.15 &   18.20 & -1.34 &  CfA  \\ 
52010.7163  &	2.90   & +3.10&   -54.90  & -1.44 &   16.70 & -0.93 &  CfA  \\
52011.6707  & -103.80  & +0.73&    53.80  & -0.56 &   16.00 & +0.20 &  CfA  \\
52032.6531  &  -27.90  & -2.23&    42.80  & +2.74 &  -49.90 & -1.49 &  CfA  \\
52034.6525  &  -37.30  & +3.87&    59.70  & -3.39 &  -55.00 & +0.49 &  CfA  \\
52038.6589  &	 2.00  & -3.19&    29.50  & +3.68 &  -65.60 & -0.49 &  CfA  \\
52211.9704  &	40.90  & -3.20&   -23.10  & +2.00 &  -60.20 & -1.13 &  CfA  \\
52237.9679  &  -69.10  & -1.98&     2.90  & +3.50 &   25.60 & -0.13 &  CfA  \\
52242.0083  &	41.30  & +0.20&  -108.00  & -1.13 &   21.30 & -0.97 &  CfA  \\
52244.9651  &	42.90  & +1.03&  -103.00  & -0.12 &   17.10 & -0.61 &  CfA  \\
52270.9681  &	79.60  & -2.61&   -61.90  & +1.64 &  -61.80 & -2.67 &  CfA  \\
52271.8541  &	21.30  & +3.59&     3.90  & -0.69 &  -61.80 & -0.16 &  CfA  \\
52273.8364  &	78.30  & -2.53&   -52.50  & +2.74 &  -65.40 & +0.30 &  CfA  \\
52274.8196  &	17.20  & +2.13&    11.50  & -0.99 &  -67.40 & -0.78 &  CfA  \\
52277.8443  &	 0.80  & +3.08&    22.00  & -3.36 &  -61.80 & +0.22 &  CfA  \\
52278.8399  &  -53.70  & -4.45&    69.40  & +1.02 &  -55.30 & +2.23 &  CfA  \\
52297.8375  &	32.40  & +0.45&  -100.50  & -0.09 &   24.80 & -0.76 &  CfA  \\
52298.8232  & -103.80  & +1.59&    40.10  & +0.24 &   25.40 & +0.59 &  CfA  \\
\hline
\end{tabular}
\end{table*}

\setcounter{table}{1}
\begin{table*}
\caption{Continued }
\begin{tabular}{@{}lrrrrrrl}
\hline
BJD & $\mathrm{RV}_A$  & $\Delta\mathrm{RV}_A$ & $\mathrm{RV}_B$  & $\Delta\mathrm{RV}_B$ & $\mathrm{RV}_C$  & $\Delta\mathrm{RV}_C$ & instr. \\ 
$-2\,400\,000$ & \multicolumn{2}{r}{(km\,s$^{-1})$} & \multicolumn{2}{c}{(km\,s$^{-1})$} & \multicolumn{2}{c}{(km\,s$^{-1})$} &  \\ 
\hline
52331.7892  &  -25.80  & -2.75  &    50.70  & +1.72   &  -63.50 & -0.01 &  CfA  \\
52332.8058  &	90.30  & -1.07  &   -66.00  & -0.61   &  -66.00 & -0.78 &  CfA  \\
52334.7602  &  -19.20  & -3.05  &    46.30  & +1.69   &  -66.20 & -0.25 &  CfA  \\
52335.6767  &	91.00  & -0.83  &   -64.40  & +1.92   &  -66.90 & -2.20 &  CfA  \\
52336.6512  &  -23.70  & +1.91  &    47.90  & -2.33   &  -61.40 & +0.68 &  CfA  \\
52337.8307  &	-1.20  & -2.44  &    20.40  & +2.82   &  -56.90 & -0.01 &  CfA  \\
52338.7667  &	80.50  & -0.50  &   -70.60  & -1.41   &  -50.00 & +1.32 &  CfA  \\
54188.7421  &	 3.40  & -1.71  &   -51.20  & +0.90   &   21.70 & -0.20 &  CfA  \\
54216.6301  &  -19.70  & +1.37  &    48.90  & -2.27   &  -52.30 & -1.60 &  CfA  \\
54218.6682  &	96.60  & +1.23  &   -66.00  & -0.52   &  -51.90 & +0.24 &  CfA  \\
54221.6492  &	94.00  & +0.91  &   -64.00  & +1.30   &  -47.10 & +2.88 &  CfA  \\
54224.6592  &	84.30  & +0.23  &   -67.60  & -0.81   &  -38.20 & +1.48 &  CfA  \\
54227.6731  &	68.30  & +1.08  &   -71.20  & +0.00   &  -18.70 & +0.22 &  CfA  \\
54422.0143  &  -69.10  & +1.09  &    20.90  & -0.18   &   31.90 & +1.35 &  CfA  \\
54423.0277  &  -52.90  & -0.52  &     8.10  & +2.09   &   28.50 & +0.83 &  CfA  \\
54423.9282  &	56.30  & -1.45  &  -102.90  & +0.18   &   24.90 & -0.14 &  CfA  \\
54425.0500  &  -77.30  & +1.52  &    38.80  & -0.30   &   21.30 & -0.39 &  CfA  \\
54425.9291  &  -51.30  & +1.20  &    16.90  & +1.73   &   19.70 & +0.67 &  CfA  \\
54430.0339  &	57.90  & +0.01  &   -84.00  & -0.18   &    6.40 & -0.07 &  CfA  \\
54456.9898  &	 3.00  & -0.01  &    28.70  & +0.13   &  -46.40 & +1.37 &  CfA  \\
54457.9095  &  -52.50  & -0.28  &    82.30  & -0.35   &  -47.50 & -1.63 &  CfA  \\
54458.8985  &	83.60  & +0.12  &   -56.70  & +1.41   &  -40.20 & +2.78 &  CfA  \\
54459.9317  &	-2.70  & +0.16  &    24.20  & -1.04   &  -38.00 & +0.85 &  CfA  \\
54461.9290  &	80.70  & -0.76  &   -71.90  & +0.28   &  -28.40 & -0.86 &  CfA  \\
54462.8983  &  -15.90  & +1.76  &    18.00  & -3.04   &  -19.40 & +1.00 &  CfA  \\
54481.8592  &  -23.80  & +0.52  &   -23.70  & -1.10   &   28.10 & +0.18 &  CfA  \\
54482.9574  &	38.10  & +2.55  &  -79.80   & +0.21   &   25.60 & +0.91 &  CfA  \\
54483.8767  &  -92.50  & +1.27  &   54.80   & +0.61   &   21.90 & -0.00 &  CfA  \\
54484.8492  &  -10.40  & -0.45  &  -27.30   & +0.53   &   18.20 & -0.72 &  CfA  \\  
54485.8154  &	49.20  & +1.02  &  -84.50   & -0.74   &   16.30 & +0.33 &  CfA  \\
54486.9101  &  -92.90  & +0.54  &   61.90   & -1.66   &   12.60 & +0.01 &  CfA  \\
54513.8629  &  -24.10  & -0.76  &   58.30   & +1.37   &  -49.60 & -0.07 &  CfA  \\
54518.8041  &  -36.90  & -0.70  &   58.60   & -0.24   &  -36.90 & +1.98 &  CfA  \\
54520.8225  &	80.70  & -2.24  &  -75.70   & -1.72   &  -27.70 & -0.15 &  CfA  \\   
54545.8346  &  -88.60  & -0.51  &   59.20   & +1.26   &   11.60 & -0.57 &  CfA  \\   
54546.6854  &	32.10  & -0.51  &  -61.60   & +0.27   &    7.60 & -1.92 &  CfA  \\   
54550.7700  &	-3.10  & +3.16  &  -11.10   & -1.64   &   -3.10 & -0.14 &  CfA  \\
54573.6194  &	93.80  & +0.26  &  -64.00   & -0.92   &  -48.70 & +0.48 &  CfA  \\
54578.7001  &	16.90  & -0.89  &   -1.80   & +0.26   &  -33.40 & +0.64 &  CfA  \\
54579.6250  &	74.00  & -0.00  &  -66.40   & -1.39   &  -30.00 & -1.57 &  CfA  \\
54968.6548  &  -17.118 & -0.967 &  6.827    & +1.334  & -28.088 & -1.064 & TRES  \\
56704.8416  &  -45.116 & +1.620 & 11.782    & +1.546  &  -1.342 & +0.481 & TRES  \\
56705.8337  &	55.089 & +0.054 & -100.501  & +0.486  &   5.987 & +0.218 & TRES  \\
56706.7765  &  -72.918 & +0.479 & 18.370    & -3.383  &  11.915 & -1.241 & TRES  \\
56708.6662  &	46.884 & +0.762 & -113.727  & +0.511  &  27.562 & +0.315 & TRES  \\
56709.8773  & -102.803 & +0.541 & ...	    &   ...   &  33.533 & -1.143 & TRES  \\
57799.3637  &	58.296 & -0.378 &  -56.802  & -3.732  & -44.185 & -2.694 & VUES  \\
58107.8212  &	76.968 & -0.111 &  -76.082  & +0.229  & -40.637 & -0.139 & UVES  \\
58107.8251  &	76.893 & +0.006 &  -75.790  & +0.332  & -40.485 & +0.006 & UVES  \\
58107.8291  &	76.695 & +0.012 &  -75.599  & +0.323  & -40.551 & -0.066 & UVES  \\
58107.8336  &	76.348 & -0.098 &  -75.448  & +0.243  & -40.603 & -0.126 & UVES  \\
58107.8376  &	76.228 & -0.002 &  -75.197  & +0.282  & -40.581 & -0.111 & UVES  \\
58107.8415  &	76.029 & +0.014 &  -75.023  & +0.243  & -40.513 & -0.050 & UVES  \\
58107.8462  &	75.898 & +0.150 &  -74.485  & +0.518  & -40.358 & +0.096 & UVES  \\
58116.7032  &	56.575 & -0.424 &  -85.526  & -0.113  & -12.247 & -0.282 & UVES  \\
\hline
\end{tabular}
\end{table*}

\setcounter{table}{1}
\begin{table*}
\caption{Continued }
\begin{tabular}{@{}lrrrrrrl}
\hline
BJD & $\mathrm{RV}_A$  & $\Delta\mathrm{RV}_A$ & $\mathrm{RV}_B$  & $\Delta\mathrm{RV}_B$ & $\mathrm{RV}_C$  & $\Delta\mathrm{RV}_C$ & instr. \\ 
$-2\,400\,000$ & \multicolumn{2}{r}{(km\,s$^{-1})$} & \multicolumn{2}{c}{(km\,s$^{-1})$} & \multicolumn{2}{c}{(km\,s$^{-1})$} &  \\ 
\hline
58126.4989  & -111.622 & +0.398 &   29.312  & +1.119  &  44.053  & +0.099 & VUES  \\
58182.3530  & -112.058 & -0.120 &   43.677  & -0.809  &  28.887  & -1.194 & VUES  \\
58184.3302  &   13.689 & +0.176 &  -93.940  & -0.083  &  40.624  & -0.036 & VUES  \\
58217.3253  &  -50.538 & +3.945 &   65.509  & -0.898  & -45.330  & +0.272 & VUES  \\
58221.3042  &    2.615 & +1.835 & ...	    & ...     & -44.353  & -0.241 & VUES  \\
58222.2875  &   74.486 & -0.741 &  -67.480  & +0.074  & -43.692  & -0.443 & VUES  \\
58222.3162  &   72.634 & -0.267 &  -65.650  & -0.429  & -43.571  & -0.351 & VUES  \\
58222.3445  &   69.667 & -0.680 &  -62.825  & -0.170  & -43.815  & -0.623 & VUES  \\
58222.3729  &   67.198 & -0.335 &  -59.450  & +0.376  & -43.773  & -0.609 & VUES  \\
58443.8394  &  -53.674 & +0.516 &   64.815  & -0.293  & -35.167  & +0.043 & VUES  \\
58467.8078  &   15.795 & +0.418 &  -25.682  & -1.106  & -15.481  & +0.522 & CHIRON  \\
58468.8109  &   47.319 & -0.392 &  -61.295  & +0.388  & -11.990  & -0.126 & CHIRON  \\
58480.7873  &  -19.796 & -0.214 &  -57.005  & +0.641  &  50.977  & +0.724 & CHIRON  \\
58494.7661  &   71.759 & +0.063 & -84.678   & +0.105  & -12.084  & +0.009 & CHIRON  \\
58508.7648  &   12.523 & +1.703 &   6.620   & -1.002  & -40.437  & +0.498 & CHIRON  \\
58526.6719  &   46.943 & +0.654 & -54.765   & -0.217  & -15.882  & -0.663 & CHIRON  \\
58541.6351  &   38.625 & +0.142 & -114.347  & +0.389  &  51.417  & +0.805 & CHIRON  \\
\hline
\end{tabular}
\end{table*}

\label{lastpage}

\end{document}